\documentclass[sigconf,balance=false]{acmart}
\setcopyright{none}
\renewcommand\footnotetextcopyrightpermission[1]{}

\usepackage{amsmath,amssymb,amsfonts}

\usepackage{graphicx}
\usepackage{subcaption} 

\usepackage{multirow}
\usepackage{booktabs}

\usepackage{float}

\usepackage{tikz}
\usetikzlibrary{shapes.geometric,arrows.meta}

\usepackage{algorithm}
\usepackage{algorithmic}

\usepackage{comment}
\usepackage{placeins}

\usepackage{rotating}   
\usepackage{pdflscape}   
\usepackage{adjustbox}   

\usepackage{xcolor} 

\acmYear{YYYY}
\acmVolume{YYYY}
\acmNumber{X}
\acmDOI{XXXXXXX.XXXXXXX}
\acmISBN{}
\acmConference{Proceedings on Privacy Enhancing Technologies}
\begin{document}
\pagestyle{plain}
\title{Dynamic Probabilistic Noise Injection for Membership Inference Defense}


\author{Javad Forough}
\orcid{0003-3399-2440}
\affiliation{%
  \institution{Imperial College London}
  \city{London}
  \country{UK}}
\email{j.forough@imperial.ac.uk}

\author{Hamed Haddadi}
\orcid{0002-5895-8903}
\affiliation{%
  \institution{Imperial College London}
  \city{London}
  \country{UK}}
\email{h.haddadi@imperial.ac.uk}


\renewcommand{\shortauthors}{Forough et al.}

\begin{abstract}
Membership Inference Attacks (MIAs) expose privacy risks by determining whether a specific sample was part of a model’s training set. These threats are especially serious in sensitive domains such as healthcare and finance. Traditional mitigation techniques, such as static differential privacy, rely on injecting a fixed amount of noise during training or inference. However, this often leads to a detrimental trade-off: the noise may be insufficient to counter sophisticated attacks or, when increased, can substantially degrade model accuracy. To address this limitation, we propose DynaNoise, an adaptive inference-time defense that modulates injected noise based on per-query sensitivity. DynaNoise estimates risk using measures such as Shannon entropy and scales the noise variance accordingly, followed by a smoothing step that re-normalizes the perturbed outputs to preserve predictive utility. We further introduce MIDPUT (Membership Inference Defense Privacy-Utility Trade-off), a scalar metric that captures both privacy gains and accuracy retention. Our evaluation on several benchmark datasets demonstrates that DynaNoise substantially lowers attack success rates while maintaining competitive accuracy, achieving strong overall MIDPUT scores compared to state-of-the-art defenses.

\end{abstract}

\keywords{membership inference attack, adaptive noise injection, 
privacy-preserving machine learning.}

\maketitle

\section{Introduction}
Machine learning has revolutionized many domains by leveraging vast amounts of data to achieve impressive performance \cite{islam2024comprehensive, forough2023anomaly, rayed2024deep, zhang2019deep, forough2022dela, sahu2023overview}. However, this success comes at a cost, where sensitive information from training datasets may be inadvertently memorized, posing serious privacy risks \cite{schwarzschild2024rethinking, el2024preserving, yang2024unveiling, ye2024privacy, li2024membership}. For example, in a scenario where a hospital deploys a predictive model to diagnose diseases; if an attacker can determine whether a patient's record was part of the training set, it may reveal that the patient has visited the hospital, thereby compromising their privacy. Similarly, in finance, membership leakage could expose clients’ investment histories. Such risks underscore the need for robust defenses against privacy attacks.

\begin{figure}[ht]
    \centering
    \includegraphics[width=1.0\columnwidth]{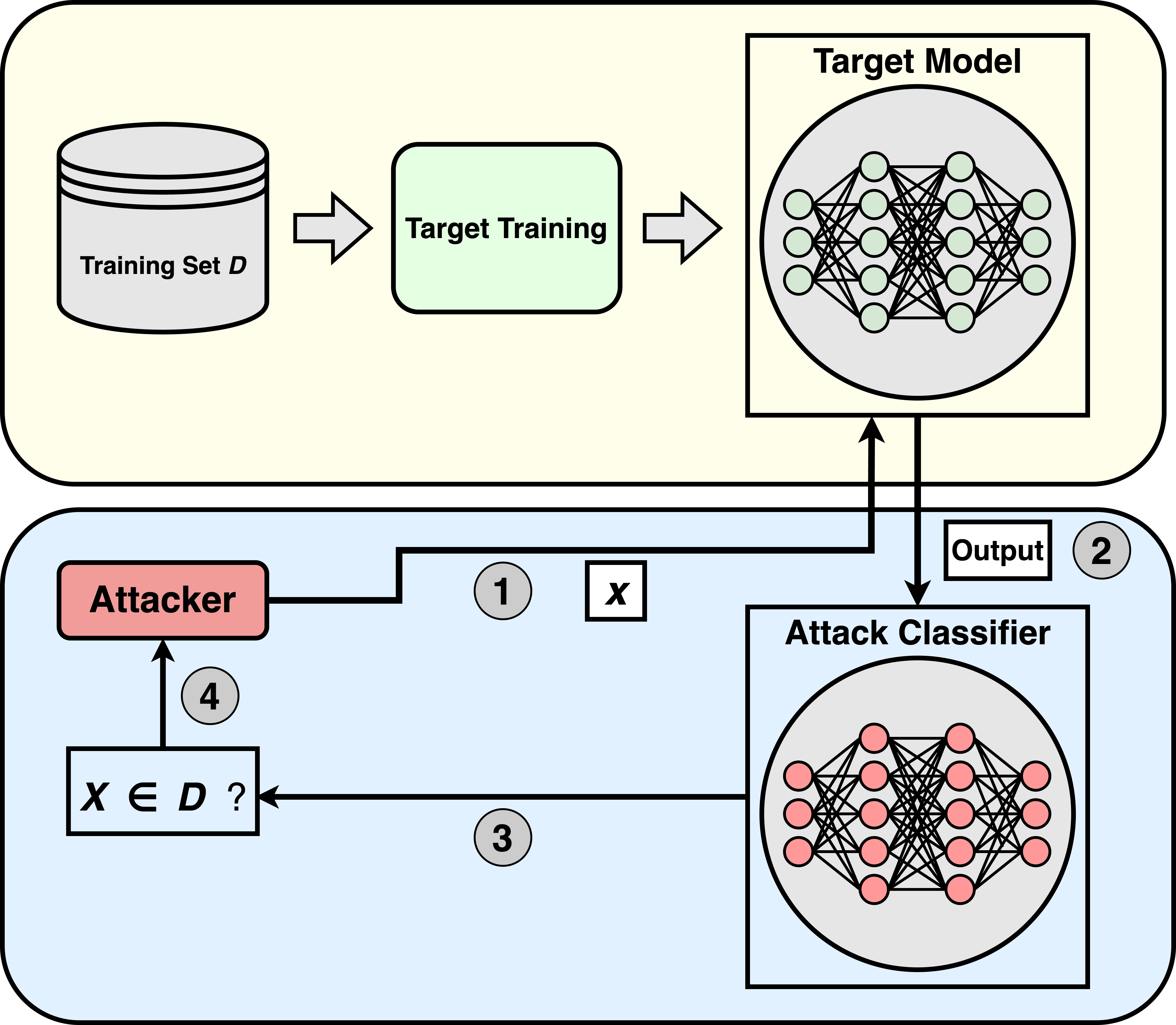} 
    \caption{Illustration of the membership inference attack (MIA) process.}
    \label{fig:MIA}
\end{figure}

Membership Inference Attacks (MIAs), as shown in Figure \ref{fig:MIA}, exploit subtle differences in a model’s output behavior to determine whether a specific data record was used during training, thereby threatening the confidentiality of individual data points \cite{shokri2017membership}. Conventional defenses, such as differential privacy, introduce a fixed level of noise during training or inference to obscure membership information \cite{abadi2016deep}. While these approaches offer formal privacy guarantees, they force an inherent trade-off between privacy and utility. In many practical settings, uniformly adding noise can significantly degrade model performance, yet reducing the noise level may leave the model susceptible to advanced membership inference attacks.

To address these challenges, we introduce DynaNoise, an adaptive noise injection approach that modulates the privacy noise dynamically based on query sensitivity. Our method first assesses the risk of each query through sensitivity analysis, by utilizing metrics such as Shannon entropy \cite{sepulveda2024applications}, and then adjusts the noise variance accordingly. A subsequent probabilistic smoothing step is applied to re-normalize the perturbed outputs, ensuring that the model retains high predictive accuracy while effectively obfuscating membership signals. Another key novelty of our approach is the introduction of a new empirical metric, the \emph{Membership Inference Defense Privacy-Utility Trade-off (MIDPUT)}, which quantitatively captures the balance between the reduction in attack success rates and the preservation of model accuracy. Our experimental results on CIFAR-10, ImageNet-10, and SST-2 datasets demonstrate that DynaNoise not only significantly reduces success rates of membership inference attacks but also achieves notably higher improvement in the MIDPUT metric compared to existing defenses.

The main contributions of this paper are threefold:
\begin{enumerate}
    \item Proposal of a novel post-hoc defense mechanism against membership inference attacks called DynaNoise that dynamically adjusts the noise level based on query sensitivity, thereby providing stronger privacy protection with minimal impact on target model accuracy.
    \item Proposal of a new empirical metric called MIDPUT, that quantifies the trade-off between the reduction in attack success rates and the loss in model performance. This metric enables a more detailed and precise evaluation of privacy defenses.
    \item Comprehensive experimental study on multiple benchmarks (CIFAR-10, ImageNet-10, and SST-2), systematically varying DynaNoise parameters and benchmarking against state-of-the-art defenses (AdvReg, MemGuard, RelaxLoss, SELENA, and HAMP). We also evaluate robustness under strong membership inference attacks, including LiRA, ensuring a fair and rigorous assessment.

\end{enumerate}

The remainder of this paper is organized as follows. In Section~\ref{sec:related_work}, we review the state-of-the-art in membership inference attacks and defenses. Section~\ref{sec:problem_statement} and ~\ref{sec:threat_model} explain our problem statement and threat model, respectively. Section~\ref{sec:background} reviews the preliminary concepts related to this work. Section~\ref{sec:proposed_approach} details the design and implementation of DynaNoise, including our adaptive noise injection and the MIDPUT metric. Section~\ref{sec:evaluation} presents our experimental setup, results, and a discussion of the advantages and limitations of the proposed approach. Finally, Section~\ref{sec:conclusion} concludes the paper and discusses future research directions.

\section{Related Work}
\label{sec:related_work}
\subsection{Membership Inference Attacks (MIAs)}

Membership inference attacks can be broadly categorized into two types: shadow training-based attacks \cite{shokri2017membership, salem2018ml, long2020pragmatic} and metric-based attacks \cite{ yeom2018privacy, song2021systematic}.

\textbf{Shadow training-based attacks.}
Shokri et al.~\cite{shokri2017membership} introduced shadow-model MIAs, where an adversary trains auxiliary models on data from a similar distribution and uses their outputs to train an attack classifier distinguishing members from non-members. Salem et al.~\cite{salem2018ml} showed that this approach can be effective even with a single shadow model, substantially reducing the attacker’s cost. Long et al.~\cite{long2020pragmatic} further improved practicality by refining shadow-model construction and the attack classifier to reduce query requirements. Overall, shadow-model attacks remain strong but typically rely on access to representative auxiliary data from the target distribution.

\textbf{Metric-based attacks.}
Metric-based attacks infer membership directly from the target model’s outputs by evaluating simple statistics, without training auxiliary models. Yeom et al.~\cite{yeom2018privacy} propose a loss-based attack that exploits the tendency of models to incur lower loss on training samples, inferring membership via thresholding. Song et al.~\cite{song2021systematic} similarly introduce confidence-based attacks that predict membership based on whether the maximum confidence exceeds a predefined threshold. While lightweight and effective, these methods are sensitive to threshold selection and model calibration.

Song and Mittal~\cite{song2021systematic} further extend this line of work by proposing entropy-based and modified entropy (M-Entropy) attacks, which exploit the observation that training samples typically produce lower-entropy and more structured output distributions. The entropy attack infers membership by thresholding Shannon entropy, while M-Entropy incorporates the probability assigned to the true label to improve robustness across confidence profiles. These attacks operate solely on output probabilities, making them well suited for black-box settings.

\textbf{Likelihood Ratio Attack (LiRA).} 
Carlini et al. \cite{carlini2022membership} recently introduced LiRA, which reframes membership inference as a statistical hypothesis test. Unlike earlier shadow-model or threshold-based approaches, LiRA estimates the distribution of model outputs on each example when it is included in training versus excluded. By fitting Gaussian distributions to these per-example losses using shadow models, LiRA computes a likelihood ratio to decide membership. This per-example calibration makes the attack far more effective at extremely low false-positive rates. Empirically, LiRA achieves up to 10x higher true-positive rates than prior attacks while also dominating in traditional aggregate metrics, establishing it as one of the strongest known MIAs.

\subsection{Membership Inference Defenses}
\label{sec:MIA-Defense}
There are several recent works aimed at addressing MIAs, which are presented and compared in Table~\ref{tab:comparative_defenses}. Subsequently, we provide a detailed overview of each method along with a discussion of their respective limitations.

Abadi et al.~\cite{abadi2016deep} propose DP-SGD, which enforces differential privacy during training by clipping per-example gradients and adding calibrated Gaussian noise, with a moments accountant used to track cumulative privacy loss. This provides formal, provable privacy guarantees at the model level. However, DP-SGD applies a static noise injection scheme that does not adapt to the sensitivity of individual queries, and the large noise required to satisfy strict privacy budgets often leads to significant accuracy degradation, particularly for complex or high-dimensional models.

Nasr et al.~\cite{nasr2018machine} introduce adversarial regularization, a training-time defense that formulates membership inference mitigation as a min--max game between a classifier and an auxiliary attack model. By jointly training the classifier to minimize attack success, this approach can reduce membership leakage with modest utility loss, but it provides no formal privacy guarantees and is sensitive to hyperparameter tuning and dataset characteristics. In a related but complementary direction, Chen et al.~\cite{chen2022relaxloss} propose RelaxLoss, which mitigates membership inference by reducing the train--test loss gap through loss relaxation and posterior flattening during training. By smoothing overconfident predictions, RelaxLoss effectively lowers attack accuracy while largely preserving utility; however, it operates solely at training time and requires retraining the model from scratch, limiting its applicability as a post-hoc defense.

Moving from modifications in the training procedure, Jia et al. \cite{jia2019memguard} propose MemGuard, a post-processing technique that directly perturbs the output confidence scores using adversarial noise. By transforming these outputs into adversarial examples, MemGuard aims to confuse any attack model attempting to distinguish between training and non-training samples. This method maintains high utility since the perturbations are designed to minimally affect the predicted labels. However, its reliance on fixed perturbation patterns means that it may be less effective if an adversary adapts to the specific noise pattern used.

Finally, Tang et al. \cite{tang2022mitigating} take an ensemble-based approach with SELENA, which trains multiple sub-models on overlapping subsets of the training data and then distills their outputs into a single prediction through a self-distillation process. This adaptive inference strategy selectively aggregates predictions from sub-models that have not seen the queried sample, thereby enhancing the privacy-utility trade-off. Despite its advantages, the ensemble inference process introduces moderate computational overhead due to the requirement of running multiple sub-models concurrently in the training phase, which can be a drawback in resource-constrained settings.

In a complementary line of work, Chen et al.~\cite{chen2024overconfidence} propose HAMP,
which mitigates membership inference by reducing model overconfidence. HAMP achieves this
through a training framework that employs high-entropy soft labels and an entropy-based
regularizer to discourage overly confident predictions. In addition, HAMP applies a
testing-time output modification that uniformly converts prediction scores into low-confidence
distributions while preserving the predicted label. As a result, confidence suppression is
applied in a query-independent manner at inference time. While this approach can effectively
reduce attack success rates, its uniform suppression may unnecessarily perturb low-risk
predictions, leading to noticeable degradation in target model accuracy, particularly under
strong membership inference attacks.

Despite these promising approaches, there are still important challenges that limit their practical effectiveness. One key limitation is that the fixed nature of noise injection does not account for the varying sensitivity of different queries. This rigidity can result in excessive noise for low-risk queries, unnecessarily degrading utility, or insufficient noise for high-risk queries, failing to adequately obfuscate membership information. Additionally, ensemble-based defenses, such as SELENA, incur significant computational and time overhead due to the need to train and evaluate multiple sub-models. These requirements may not be feasible in resource-constrained environments. To address these limitations, our work proposes a dynamic noise injection mechanism that adjusts the noise level based on query sensitivity, thereby achieving a more balanced trade-off between privacy protection and model performance, while incurring only negligible computational overhead.

\textbf{Novelty compared to prior noise-based defenses:}
Prior defenses against membership inference attacks have perturbed model outputs with
random noise, either during training or at inference time, typically using constant magnitudes
or coarse, confidence-based heuristics, such as relying solely on the highest predicted
probability, to set the noise level. While such approaches can reduce attack success rates,
they often fail to fully leverage the structure of the entire output distribution and may apply
suboptimal noise amounts across queries. 
In addition, defenses such as HAMP~\cite{chen2024overconfidence} mitigate
membership leakage by reducing overconfidence and applying a uniform, query-independent
confidence suppression at inference time, without adapting the level of protection to the
uncertainty or risk of individual queries.

DynaNoise departs from these patterns in several key ways. First, rather than basing noise
magnitude solely on the highest predicted probability or applying uniform suppression, we
derive a query-specific sensitivity score from the Shannon entropy of the full logit
distribution, capturing the model's overall uncertainty. Second, we integrate Gaussian
perturbation with a principled temperature-scaling step to re-normalize the perturbed logits,
ensuring controlled smoothing of the outputs. Third, our adaptation is continuous and
fine-grained, with the noise level varying smoothly across queries instead of being assigned
via thresholds or bins. Finally, DynaNoise is applied purely as a post-hoc mechanism,
requiring no model retraining or architectural modifications, making it lightweight and
easily deployable in practice.

\begin{table*}[ht]
\centering
\Large
\caption{Comparison of defense mechanisms against membership inference attacks.}
\resizebox{\textwidth}{!}{%
\begin{tabular}{|l|l|p{4.0cm}|l|l|l|}
\hline
\textbf{Defense Method} & \textbf{Year} & \textbf{Approach} & \textbf{Privacy Guarantee} & \textbf{Utility Impact} & \textbf{Comp. Overhead} \\ \hline
DP-SGD \cite{abadi2016deep} & 2016 & Gradient clipping + additive Gaussian noise (static noise injection) & Formal (provable DP) & High accuracy drop & Moderate \\ \hline
Adversarial Regularization \cite{nasr2018machine} & 2018 & Min-max adversarial training integrating an attack model into training & Empirical & Low to moderate accuracy drop & Moderate \\ \hline
MemGuard \cite{jia2019memguard} & 2019 & Post-processing: adversarial noise added to confidence score vectors & Empirical & Minimal accuracy drop & High (Inference) \\ \hline
RelaxLoss \cite{chen2022relaxloss} & 2022 & Loss-relaxation with posterior flattening and alternating ascent/descent to reduce train--test loss gap (confidence smoothing) & Empirical & Low to moderate accuracy drop & Low \\ \hline
SELENA \cite{tang2022mitigating} & 2022 & Adaptive ensemble (Split-AI + Self-Distillation) with dynamic noise injection & Empirical & Minimal accuracy drop & Moderate \\ \hline

HAMP \cite{chen2024overconfidence} & 2024 & Training-time overconfidence mitigation with entropy regularization and high-entropy soft labels, combined with uniform confidence suppression at inference time & Empirical & Moderate to high accuracy drop & Low \\ \hline

DynaNoise (This work) & 2026 & Adaptive noise injection based on query sensitivity (sensitivity analysis, dynamic noise variance modulation, probabilistic smoothing) & Empirical & Minimal accuracy drop & Low \\ \hline
\end{tabular}
}
\label{tab:comparative_defenses}
\end{table*}

\section{Problem Statement}
\label{sec:problem_statement}

Let \( f_\theta : \mathcal{X} \rightarrow \mathcal{Y} \) denote a trained model parameterized by \(\theta\), which outputs a probability distribution \( f_\theta(q) = (p_1, p_2, \ldots, p_k) \) over \(k\) classes for an input sample \( q \in \mathcal{X} \). Membership inference attacks aim to determine whether a given query \(q\) was part of the model’s training dataset \(D_{\text{train}}\). Formally, an adversary attempts to infer the membership status of \(q\) by distinguishing between two hypotheses:
\[
H_0: q \notin D_{\text{train}}, \qquad H_1: q \in D_{\text{train}}.
\]
Such attacks exploit the distributional discrepancies between the model’s output probabilities for members versus non-members, i.e., \( f_\theta(q_m) \neq f_\theta(q_n) \) for \(q_m \in D_{\text{train}}\) and \(q_n \notin D_{\text{train}}\).

Traditional privacy-preserving mechanisms, such as differential privacy (DP), mitigate this risk by adding a fixed amount of random noise $\eta$ to the model outputs:
\[
\tilde{f}_\theta(q) = f_\theta(q) + \eta, \quad \eta \sim \mathcal{N}(0, \sigma^2 I),
\]
where \(\sigma\) is a pre-defined noise scale. However, this uniform perturbation fails to consider the heterogeneous sensitivity of queries. For low-risk queries, excessive noise can unnecessarily degrade prediction accuracy, while for high-risk queries, insufficient noise may still leak membership information.

The key challenge, therefore, is to design a mechanism that achieves an optimal balance between privacy and utility by \emph{adapting} the injected noise to the sensitivity of each query. Specifically, the goal is to learn or define a sensitivity function \(R(q)\) such that the noise variance \(\sigma^2(q)\) scales with the privacy risk of the query, thereby maximizing protection for sensitive cases without impairing overall model performance.

\section{Threat Model}
\label{sec:threat_model}

In our threat model, the adversary is granted \textit{black-box access} to the target model. Specifically, the attacker can submit any input query \(q\) and obtain its corresponding prediction vector, i.e., the post-softmax probability output $\hat{f}(q)$. The adversary is assumed to know the input/output format (e.g., number of classes and range of confidence values) and may either (i) possess partial knowledge of the model architecture and training procedure, or (ii) interact with the model through a machine-learning-as-a-service (MLaaS) interface where internal parameters remain hidden. We assume that the adversary has access to the full prediction vector (e.g.,
confidence or probability scores). As a result, DynaNoise is designed to defend
against score-based membership inference attacks and does not address attacks
that rely solely on predicted labels.

The adversary’s objective is to determine whether a given record \(q\) was used in training the target model. Formally, the attack attempts to distinguish between:
\[
H_0: \; q \notin D_{\text{train}}, \qquad
H_1: \; q \in D_{\text{train}},
\]
where \(D_{\text{train}}\) denotes the model’s training dataset. The adversary may also have access to auxiliary information about the underlying data distribution, such as population-level feature statistics. An attack is considered successful if it can infer the membership status of \(q\) with accuracy significantly better than random guessing.

\noindent\textbf{Query assumptions:}
DynaNoise operates under a \emph{limited-query} assumption, where repeated identical queries are not allowed or are infrequent. This reflects common practical scenarios in deployed MLaaS systems that log or rate-limit queries to prevent excessive access. If the same query were submitted multiple times, an adversary could theoretically average the noisy outputs and approximate the unperturbed logits. Addressing such adaptive query repetition is outside the current scope but can be mitigated by caching consistent noisy responses or incorporating query-count--dependent noise scaling, as discussed in Section~\ref{sec:limitations}.
  
Under these conditions, DynaNoise dynamically adjusts the noise added to each output based on the sensitivity of the query, thereby reducing the adversary’s ability to distinguish between members and non-members while preserving model utility.

\section{Preliminaries}
\label{sec:background}
In this section, we explain the background concepts related to this work. In addition, Table~\ref{tab:notations} presents a summary of the notations used throughout this paper.

\subsection{Static Differential Privacy and Its Limitations}
Static differential privacy (DP) is a formal framework for protecting individual data records by adding random noise during model training or inference \cite{abadi2016deep, jayaraman2019evaluating}. The goal is to ensure that the inclusion or exclusion of any single data point has only a limited impact on the model’s output, thus protecting individual privacy.

A randomized mechanism \(M\) (such as a learning algorithm or model output function) is said to satisfy \((\epsilon, \delta)\)-differential privacy if, for all possible outputs \(S\), and for any pair of neighboring datasets \(D\) and \(D'\) (which differ in only one data record), the following inequality holds:
\[
\Pr[M(D) \in S] \leq e^\epsilon \cdot \Pr[M(D') \in S] + \delta,
\]
where:
\begin{itemize}
    \item \(\Pr[M(D) \in S]\): the probability that the mechanism produces an output within the set \(S\), where the randomness arises from the noise intentionally added by the mechanism.
    \item \(M(D)\): the randomized output (e.g., model prediction or learned parameters) when the mechanism operates on dataset \(D\),
    \item \(S\): any subset of possible outputs,
    \item \(D, D'\): neighboring datasets differing in exactly one entry,
    \item \(e^\epsilon\): a multiplicative bound controlling the degree of output similarity between neighboring datasets; \(e\) is Euler's number (approximately 2.718)
    \item \(\epsilon\): the privacy budget (smaller values indicate stronger privacy),
    \item \(\delta\): the probability of violating the \(\epsilon\)-bound (typically a small value like \(10^{-5}\)).
\end{itemize}

While this approach provides mathematically rigorous privacy guarantees, it usually applies the same level of noise to all data points or queries, regardless of their actual risk level. This static and uniform noise injection leads to a key limitation: 
\begin{itemize}
    \item If the noise is too large, it can significantly degrade the model's accuracy and utility.
    \item If the noise is too small, it may fail to protect against advanced membership inference attacks.
\end{itemize}

This inherent rigidity in static DP motivates the need for more flexible strategies, such as adaptive or dynamic noise injection mechanisms that tailor the amount of noise based on the sensitivity of each query or prediction.

\subsection{Information Leakage in Deep Learning}
Deep neural networks are known to exhibit complex behavior that may inadvertently reveal information about their training data. Several works \cite{xu2019ganobfuscator, pan2024differential, veetil2024analysis} have shown that even models with strong generalization capabilities can overfit on certain examples, thereby creating a gap between the output distributions for training and non-training data. This leakage is often measured using differences in loss or divergence in prediction distributions, which can be formally expressed by metrics such as Kullback-Leibler divergence:
\[
D_{KL}(P \parallel Q) = \sum_{i} P(i) \log \frac{P(i)}{Q(i)},
\]
where \(P\) and \(Q\) represent the output distributions for training and non-training examples, respectively.

Understanding and quantifying this leakage is critical for designing effective privacy-preserving mechanisms. The degree of leakage can inform the design of adaptive noise injection strategies that modulate the amount of noise based on the risk associated with each query, thereby reducing the adversary’s ability to distinguish between members and non-members while preserving the utility of the model.

\begin{table}[htb]
\centering
\caption{Summary of notations used in the paper.}
\resizebox{\columnwidth}{!}{
\begin{tabular}{c|p{5.1cm}}
\hline
\textbf{Symbol} & \textbf{Description} \\
\hline
\( f(q) \) & Model logits for input \(q\). \\

\( \tilde{f}(q) \) & Noisy logits: \(f(q) + \eta\). \\

\( \hat{f}(q) \) & Final probabilities (after smoothing). \\

\( k \) & Number of classes. \\

\( \mathbf{p} \) & Softmax of \(f(q)\). \\

\( H(p) \) & Entropy of \(\mathbf{p}\). \\

\( R(q) \) & Sensitivity score of \(q\). \\

\( \sigma_0^2 \) & Base noise variance. \\

\( \lambda \) & Noise scaling parameter. \\

\( \sigma(q)^2 \) & Adjusted noise variance. \\

\( \eta \) & Gaussian noise vector. \\

\( T \) & Temperature for smoothing. \\

\(\text{ASR}\) & Attack Success Rate. \\

\(\text{MIDPUT}_{Overall}\) & Overall Privacy--Utility Trade-off metric. \\

\(\text{MIDPUT}_{\text{Conf}}, \text{MIDPUT}_{\text{Loss}}\) & Per-attack MIDPUT metrics for Confidence-based and Loss-based attacks. \\

\(\text{MIDPUT}_{\text{Shadow}}, \text{MIDPUT}_{\text{LiRA}}\) & Per-attack MIDPUT metrics for Shadow-model and LiRA attacks. \\

\(\text{MIDPUT}_{\text{Ent}}, \text{MIDPUT}_{\text{M-Ent}}\) & Per-attack MIDPUT metrics for entropy-based and M-Entropy membership inference attacks. \\

\hline
\end{tabular}}
\label{tab:notations}
\end{table}

\section{Proposed Approach}
\label{sec:proposed_approach}

Our proposed approach dynamically mitigates membership inference attacks by adapting the noise injection process to the sensitivity of each individual query. The complete procedure is summarized in Algorithm~\ref{alg:dyna-noise}. The system is composed of three primary components, as described below.

\begin{itemize}
    \item \textbf{Sensitivity Analysis:} \\
    In this module, we evaluate the risk associated with each query based on the model's output probability distribution. Let \( p = (p_1, p_2, \ldots, p_k) \) denote the output probability vector for a given input, where \( k \) is the number of classes. We compute the Shannon entropy \cite{sepulveda2024applications}:
    \[
    H(p) = -\sum_{i=1}^{k} p_i \log(p_i),
    \]
    which quantifies the uncertainty of the model's prediction. We then define the normalized sensitivity score for query \( q \) as:
    \[
    R(q) = 1 - \frac{H(p)}{\log k},
    \]
    ensuring \( R(q) \in [0,1] \). A higher \( R(q) \) indicates that the model is highly confident (i.e., low entropy) in its prediction, and therefore at greater risk of leaking membership information.

    \item \textbf{Dynamic Noise Injection:} \\
    Using the computed sensitivity score, we dynamically adjust the noise added to the model’s output logits. The noise variance scales as:
    \[
    \sigma(q)^2 = \sigma_0^2 \left(1 + \lambda R(q)\right),
    \]
    where \(\sigma_0^2\) is the base noise variance and \(\lambda\) is a scaling parameter that amplifies the perturbation for higher-risk queries. Gaussian noise \(\eta \sim \mathcal{N}(0, \sigma(q)^2)\) is then added to the raw logits \( f(q) \), producing perturbed outputs \( \tilde{f}(q) = f(q) + \eta \). This ensures that high-risk queries receive proportionally more noise, effectively obfuscating membership signals.

    \item \textbf{Probabilistic Smoothing:} \\
    To mitigate distortions introduced by noise injection while maintaining predictive performance, we apply a probabilistic smoothing operation by re-normalizing the perturbed logits using a temperature-scaled softmax:
    \[
    \hat{f}(q) = \text{softmax}\!\left(\frac{\tilde{f}(q)}{T}\right),
    \]
    where \(T > 1\) controls the sharpness of the probability distribution and provides a trade-off between smoothing and discriminative power.
\end{itemize}

\begin{figure*}[ht]
    \centering
    \includegraphics[width=1.0\textwidth]{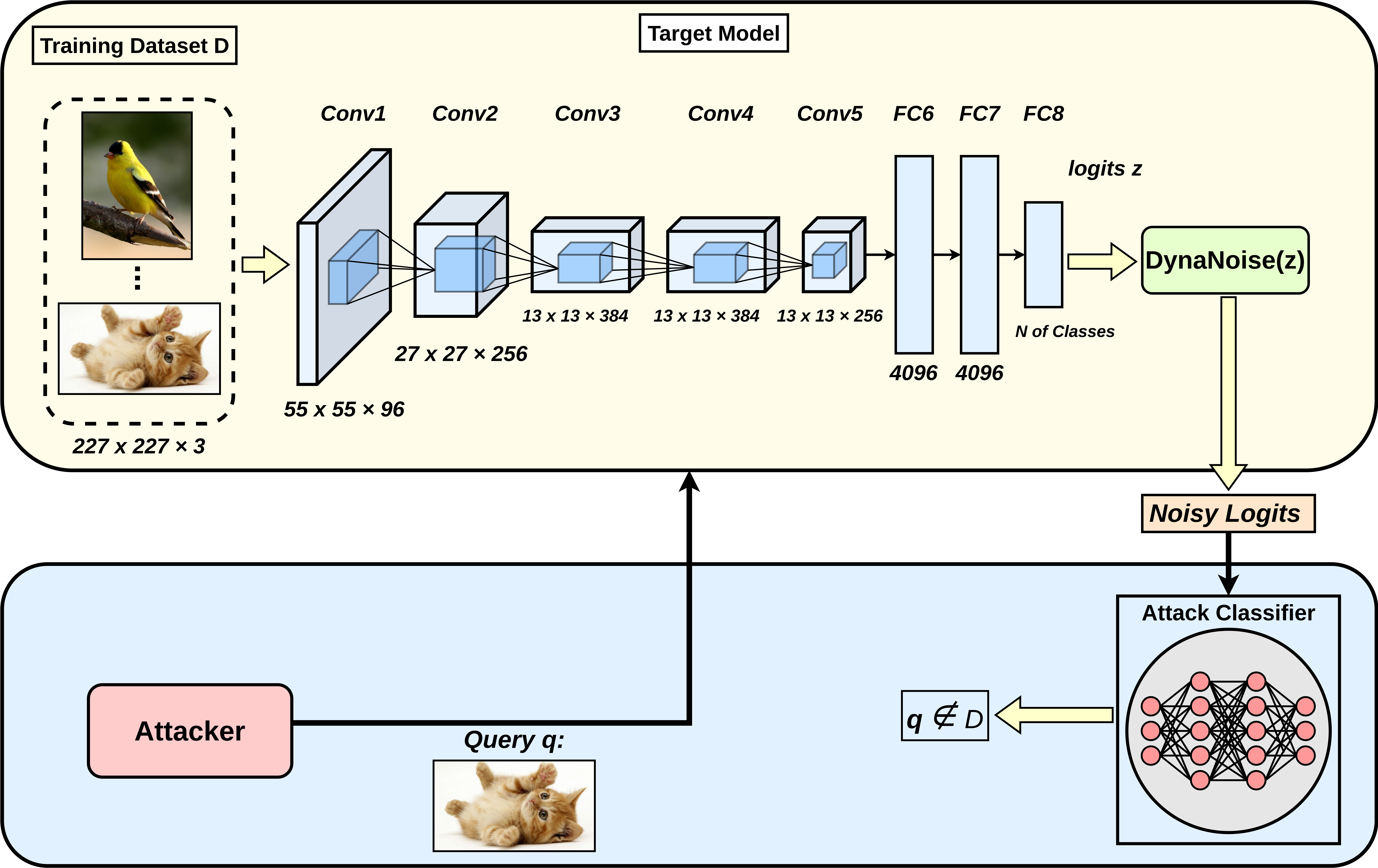}
    \caption{Overview of the proposed DynaNoise approach integrated into the AlexNet model for illustration. The noise injection process adapts to the model’s uncertainty, providing enhanced obfuscation against membership inference attacks.}
    \label{fig:DynaNoise}
\end{figure*}

By integrating sensitivity analysis, dynamic noise injection, and probabilistic smoothing, DynaNoise adapts to the context of each inference request. This design effectively obscures membership cues in high-risk queries while maintaining high accuracy and low computational overhead. Figure~\ref{fig:DynaNoise} provides an overview of the DynaNoise post-processing pipeline and its integration into the target model.

\begin{algorithm}[ht]
\caption{DynaNoise: Adaptive Noise Injection Based on Query Sensitivity}
\label{alg:dyna-noise}
\begin{algorithmic}[1]
\STATE \textbf{Input:} Trained model \( f(\cdot) \) with \( k \) output classes, base noise variance \( \sigma_0^2 \), scaling parameter \( \lambda \), temperature parameter \( T > 1 \), and input query \( q \).
\STATE \textbf{Output:} \( \hat{f}(q) \), the final probability vector after noise injection and smoothing.

\STATE \textbf{Step 1: Compute Raw Output}
    \STATE \quad Compute logits \( f(q) \in \mathbb{R}^{k} \).
    \STATE \quad Compute probability vector \( p = \text{softmax}\!\bigl(f(q)\bigr) \).

\STATE \textbf{Step 2: Sensitivity Analysis}
    \STATE \quad Compute Shannon entropy \( H(p) = -\sum_{i=1}^{k} p_i \log(p_i) \).
    \STATE \quad Compute sensitivity score \( R(q) = 1 - H(p)/\log k \).

\STATE \textbf{Step 3: Dynamic Noise Injection}
    \STATE \quad Compute variance \( \sigma(q)^2 = \sigma_0^2 \bigl(1 + \lambda R(q)\bigr) \).
    \STATE \quad Sample noise \( \boldsymbol{\eta} \sim \mathcal{N}\!\bigl(\mathbf{0}, \sigma(q)^2 I_k\bigr) \in \mathbb{R}^{k} \).
    \STATE \quad Perturb logits \( \tilde{f}(q) = f(q) + \boldsymbol{\eta} \).

\STATE \textbf{Step 4: Probabilistic Smoothing}
    \STATE \quad Compute final output \( \hat{f}(q) = \text{softmax}\!\bigl(\tilde{f}(q)/T\bigr) \).

\STATE \textbf{Return:} \( \hat{f}(q) \).
\end{algorithmic}
\end{algorithm}

Although DynaNoise is primarily an empirical defense, its adaptive noise modulation is theoretically motivated by the notion of \emph{sensitivity} in Differential Privacy (DP). Here we explain how the query sensitivity measure \(R(q)\) conceptually relates to DP and why this provides an interpretable foundation for privacy protection.
\noindent\textbf{Connection to differential privacy:}
In $(\epsilon, \delta)$-DP, the \emph{global sensitivity} of a mechanism \( f \) is defined as
\[
\Delta f = \max_{D, D'} \| f(D) - f(D') \|,
\]
where \( D \) and \( D' \) differ by a single record. The Gaussian mechanism achieves privacy by adding noise proportional to this sensitivity, ensuring that the influence of any individual record on the output remains limited.
In DynaNoise, the model’s output probability vector for a query \(q\) serves as the observable function \(f(D)\). Instead of computing dataset-level differences, DynaNoise estimates a \emph{per-query sensitivity proxy} using Shannon entropy: queries with low entropy (high confidence) are empirically linked to greater membership risk. The normalized entropy complement \(R(q)\) thus functions as a practical analogue to local sensitivity, determining the magnitude of noise required to conceal the query’s influence.
\noindent\textbf{Approximate privacy analogy:}
As defined in the previous section, DynaNoise scales the variance of the added Gaussian noise with \(R(q)\). Since in DP the privacy loss parameter $\epsilon$ is inversely proportional to the noise magnitude, DynaNoise implicitly assigns each query an \emph{effective privacy budget}:
\[
\epsilon(q) \propto \frac{1}{\sigma(q)}.
\]
Queries with higher sensitivity (low entropy) therefore receive larger perturbations, corresponding to smaller effective $\epsilon$ (stronger privacy), while low-sensitivity queries incur smaller noise to preserve utility.

\subsection{Entropy as a Proxy for Sensitivity}
\label{sec:entropy-jacobian-connection}

DynaNoise modulates output noise according to the Shannon entropy 
$H(\mathbf{p}(q))$ of the model’s predictive distribution,
\[
\mathbf{p}(q) = \text{softmax}(\mathbf{z}(q)), \quad 
\mathbf{z}(q) = f_\theta(q).
\]
While entropy measures predictive uncertainty, it also serves as a lightweight surrogate for local sensitivity. Intuitively, entropy reflects how concentrated the model’s confidence is across classes, which determines how the output responds to small input perturbations. Formally, if $J_f(q)=\frac{\partial f_\theta(q)}{\partial q}$ denotes the output Jacobian, then under a first-order approximation $f_\theta(q+\delta)\approx f_\theta(q)+J_f(q)\delta$, and the resulting change in the softmax output scales with $\|J_f(q)\|$, a local Lipschitz proxy for sensitivity~\cite{cisse2017parseval, novak2018sensitivity}. 

As shown in prior works ~\cite{ross2018improving,novak2018sensitivity,tsipras2019robustness}, low-entropy (peaked) predictions correspond to overconfident regions with sharper and more brittle decision boundaries, whereas high-entropy predictions yield smoother and more stable responses. Empirically, we observe a consistent relationship between entropy and Jacobian-based sensitivity across all evaluated models. Detailed scatter plots for CIFAR-10, ImageNet-10, and SST-2—showing Pearson correlations between $H(\mathbf{p}(q))$ and $\|J_f(q)\|_2$—are provided in Appendix~\ref{app:entropy-sensitivity-plots}. These visualizations confirm that entropy reliably captures the model's exposure level, supporting its use as a calibration-free proxy for sensitivity within DynaNoise.

\section{Evaluation}
\label{sec:evaluation}
We evaluated our approach on several benchmark datasets and model architectures that are explained in Section \ref{sec:Datasets}. Our analysis focuses on three primary aspects: (i) the effectiveness of membership inference defenses; (ii) the impact on model accuracy (i.e., utility); and (iii) the computational cost of the defense mechanism

\subsection{Metrics}
\label{sec:metrics}
The following metrics were used to evaluate the performance of our approach:
\begin{itemize}
    \item \textbf{Attack Success Rate (ASR):} The fraction of correctly inferred membership, defined as
    \[
    \text{ASR} = \frac{N_{\text{correct}}}{N_{\text{total}}},
    \]
    where \(N_{\text{correct}}\) is the number of correct membership predictions and \(N_{\text{total}}\) is the total number of predictions.
    \item \textbf{Model Accuracy:} The overall classification accuracy of the target model, measured both before and after applying a defense.
    \item \textbf{Membership Inference Defense Privacy--Utility Trade-off (MIDPUT):}
    Our proposed metric for quantifying the trade-off between privacy protection and model utility.
    Let
    \[
    \Delta_{\text{acc}}
    =
    \text{test\_acc}_{\text{no\_def}}
    -
    \text{test\_acc}_{\text{def}}
    \]
    denote the accuracy drop induced by a defense, and for each attack
    $A \in \{\text{conf}, \text{loss}, \text{ent}, \text{m-ent}, \text{shadow}, \text{lira}\}$,
    let
    \[
    \Delta_A
    =
    \text{attack\_acc}_{\text{no\_def}}^{(A)}
    -
    \text{attack\_acc}_{\text{def}}^{(A)}
    \]
    denote the corresponding reduction in attack success rate.
    
    The overall MIDPUT score is defined as
    \[
    \text{MIDPUT}_{\text{Overall}}
    =
    \frac{1}{6}
    \sum_{A}
    \Delta_A
    -
    \Delta_{\text{acc}},
    \]
    where the summation is taken over all evaluated attack types.
    The per-attack MIDPUT scores are given by
    \[
    \text{MIDPUT}_A
    =
    \Delta_A
    -
    \Delta_{\text{acc}}.
    \]
    
    Across our experiments, MIDPUT values typically lie in the range $[-1,1]$, where values closer to $1$
    indicate strong privacy gains with minimal utility loss, while values closer to $-1$ reflect
    unfavorable trade-offs dominated by accuracy degradation.

\end{itemize}

\noindent\textbf{Extensibility of MIDPUT:}
The MIDPUT metric is not tied to any specific attack and can incorporate additional attacks in a modular way. In this work, we report both per-attack MIDPUT scores (e.g., MIDPUT\(_{\text{Conf}}\), MIDPUT\(_{\text{Loss}}\), MIDPUT\(_{\text{Entropy}}\),
MIDPUT\(_{\text{M\text{-}Entropy}}\), MIDPUT\(_{\text{Shadow}}\), MIDPUT\(_{\text{LiRA}}\))
 as well as an overall MIDPUT that averages across attacks. This design serves two purposes: (i) it prevents conclusions from depending on a single attack model, and (ii) it enables future work to plug in stronger or newly proposed membership inference attacks without changing the structure of the evaluation protocol. In other words, MIDPUT is a general scoring framework for privacy--utility trade-offs rather than a metric tailored to any specific attack.

\noindent\textbf{Interpretation and role of MIDPUT:}
Traditional privacy--utility evaluations often rely on Pareto frontiers, which visually illustrate trade-offs but are difficult to compare across models or attack settings. The proposed MIDPUT metric provides a complementary scalar summary that quantifies the joint balance between privacy preservation and utility retention. By aggregating the reduction in attack success rate and the preservation of accuracy across multiple attack types, MIDPUT yields a single interpretable score that enables consistent benchmarking and ranking of defenses.

The aggregated score MIDPUT$_{\text{Overall}}$ is intended as a complementary summary measure of the privacy--utility trade-off rather than a worst-case or per-attack security guarantee. MIDPUT$_{\text{Overall}}$ aggregates per-attack improvements by averaging across multiple attack types, and therefore should not be interpreted as dominance against any specific attack. Individual attacks may exhibit substantially different behavior, and strong performance under one attack does not imply robustness under all attacks.

Rather than replacing Pareto analyses, MIDPUT complements them by offering a concise, quantitative measure of overall defense effectiveness within the privacy--utility landscape. For this reason, we report per-attack MIDPUT scores (e.g., MIDPUT$_{\text{Conf}}$, MIDPUT$_{\text{Loss}}$, MIDPUT$_{\text{Entropy}}$,
MIDPUT$_{\text{M\text{-}Entropy}}$, MIDPUT$_{\text{Shadow}}$, MIDPUT$_{\text{LiRA}}$)
 and raw attack success rates alongside MIDPUT$_{\text{Overall}}$, and emphasize that security conclusions should be drawn primarily from attack-specific results. MIDPUT$_{\text{Overall}}$ serves as a concise, high-level indicator that complements per-attack analysis and Pareto-style evaluations, rather than replacing them.

\subsection{Membership Inference Attacks}
We implemented four distinct membership inference attacks:
\begin{enumerate}
    \item \textbf{Confidence threshold attack:} Predicts a sample as a member if the maximum predicted probability exceeds a fixed threshold \(\tau\):
    \[
    \text{Predict "in" if } \max_{i} p_i > \tau.
    \]
    In our evaluation, we set \(\tau = 0.9\)
    
    \item \textbf{Loss threshold attack:} Computes the cross-entropy loss 
    \[
    \ell = -\log p(y)
    \]
    for the true label \(y\), and predicts membership if \(\ell < \gamma\), where \(\gamma\) is a fixed threshold. In our evaluation, we set \(\gamma = 0.5\).

    \textbf{Entropy-based attacks:}  
    Following Song and Mittal~\cite{song2021systematic}, we evaluate two entropy-based membership
    inference attacks that operate directly on the model’s output distribution.
    The \emph{Entropy} attack infers membership by thresholding the Shannon entropy
    \[
    H(\mathbf{p}) = -\sum_{i} p_i \log p_i,
    \]
    with lower entropy indicating higher likelihood of membership.
    The \emph{Modified Entropy (M-Entropy)} attack refines this criterion by incorporating the
    probability assigned to the true label, improving robustness across varying confidence profiles.
    Both attacks require only black-box access to predicted probabilities.

    \item \textbf{Shadow-model attack:} Trains a shadow model on an auxiliary dataset drawn from the same distribution as the target model's training dataset. From the shadow model’s outputs, features such as maximum confidence, cross-entropy loss, and confidence margin are extracted to train a binary classifier \(A(\mathbf{x})\), where \(\mathbf{x}\) is the feature vector. The membership decision is then based on the classifier’s output. In our setup, we allocate 70\% of the available data to training and evaluating the target model, while the remaining 30\% is reserved for constructing the shadow model and training the attack classifier. We use the same architecture for the shadow model as the target model to closely mimic its behavior, and we employ a logistic regression classifier as the final attack model.

        \item \textbf{Likelihood Ratio Attack (LiRA):}  
    We also evaluate the Likelihood Ratio Attack (LiRA) \cite{carlini2022membership}, a state-of-the-art membership inference attack. Unlike threshold-based or shadow-classifier approaches, LiRA formulates membership inference as a per-sample likelihood test. Concretely, we first train multiple shadow models on disjoint subsets of data sampled from the same distribution as the target model. For each shadow model, we record the per-sample cross-entropy losses on both \emph{member (IN)} examples, which are those used in training the shadow model, and \emph{non-member (OUT)} examples, those excluded from training. Gaussian distributions are then fitted to these two sets of losses, yielding estimates of \(P(\ell \mid \text{IN})\) and \(P(\ell \mid \text{OUT})\), where \(\ell\) denotes the loss.  

    For each target sample, we compute the log-likelihood ratio
    \[
    \text{LLR}(x) = \log \frac{P(\ell(x) \mid \text{IN})}{P(\ell(x) \mid \text{OUT})},
    \]
    and classify membership based on the calibrated sign of this statistic. Importantly, post-hoc defenses such as DynaNoise are applied only when evaluating the \emph{target model}, while the attacker's shadow models are always trained without defenses. In our setup, we train five shadow models per dataset (same architecture as the target), and use their pooled losses to fit the Gaussian parameters and calibrate the decision rule.

\end{enumerate}

\noindent\textbf{Rationale for attack selection:}
The attacks above were chosen to cover the major classes of membership inference strategies studied
in the literature. The confidence-threshold and loss-threshold attacks represent
\textit{metric-based} MIAs that exploit scalar statistics derived from model outputs, such as maximum
confidence or per-sample loss. Entropy-based and modified entropy attacks further extend this class
by leveraging global properties of the output distribution, capturing overconfidence patterns that
are not visible through single-score metrics. Shadow-model attacks represent the class of
\textit{supervised attacker training}, in which the adversary trains an auxiliary classifier to
explicitly learn a member versus non-member decision rule from model outputs. Finally, LiRA is a
state-of-the-art \textit{adaptive statistical attack} that performs per-sample likelihood testing
using calibrated shadow models and is known to substantially outperform earlier methods,
particularly at low false-positive rates.

Each membership inference attack is systematically evaluated under the following six defense conditions:

\begin{enumerate}
    \item \textbf{No defense:} The target model's raw outputs are directly exposed to the attacker without any privacy preserving approach applied.
    \item \textbf{Adversarial Regularization (AdvReg) \cite{nasr2018machine} Defense:} The model is trained with an auxiliary adversary that penalizes leakage of membership information during optimization. This regularization term encourages the model to produce output distributions that are indistinguishable between members and non-members, thereby reducing attack success rates.

    \item \textbf{MemGuard \cite{jia2019memguard} defense:}  
    MemGuard generates small adversarial perturbations to output probabilities at inference time, reducing membership leakage without modifying or retraining the underlying model.
    
    \item \textbf{RelaxLoss \cite{chen2022relaxloss} defense:} The RelaxLoss defense is applied during model training, where a modified loss function penalizes overconfident predictions to reduce overfitting. This adjustment smooths the model’s decision boundaries and lowers membership leakage, providing an effective privacy-utility trade-off without requiring additional noise or architectural changes.

    \item \textbf{SELENA \cite{tang2022mitigating} defense:} The SELENA defense approach is employed, which leverages ensemble learning and self-distillation to mitigate membership leakage prior to the execution of the attack.

    \item \textbf{HAMP \cite{chen2024overconfidence} defense:} 
    The HAMP defense is applied by mitigating model overconfidence through entropy-regularized training with high-entropy soft labels, followed by a testing-time output modification that uniformly suppresses prediction confidence in a query-independent manner.

    \item \textbf{DynaNoise defense:} The proposed DynaNoise approach is applied to the model outputs in a post-processing step, where it dynamically injects calibrated probabilistic noise based on query sensitivity to conceal membership signals while preserving overall model utility.
\end{enumerate}

\subsection{Datasets and Models}
\label{sec:Datasets}
Experiments were conducted on three widely used benchmark datasets that span both image and text domains. The datasets are described in detail below:

\begin{itemize}
    \item \textbf{CIFAR-10\footnote{https://www.tensorflow.org/datasets/catalog/cifar10}:} This dataset consists of 60,000 color images of size 32×32 distributed across 10 balanced classes. CIFAR-10 is a standard benchmark in computer vision, widely used to evaluate image classification models under moderate complexity conditions. Its relatively low resolution and balanced class distribution make it ideal for testing both model performance and the effectiveness of privacy-preserving techniques.
    \item \textbf{ImageNet-10\footnote{https://www.kaggle.com/datasets/liusha249/imagenet10/code}:} A curated subset of the larger ImageNet dataset, ImageNet-10 contains images from 10 diverse classes. This subset is more challenging than CIFAR-10 due to its greater variability in image content, higher resolution, and increased intra-class diversity. It provides a rigorous testbed for evaluating the scalability and robustness of defense mechanisms on complex, real-world data.
    \item \textbf{SST-2\footnote{https://huggingface.co/datasets/gimmaru/glue-sst2}:} The Stanford Sentiment Treebank (SST-2) is a sentiment classification dataset extracted from the GLUE benchmark. It comprises text samples labeled with binary sentiment (positive or negative). SST-2 is representative of natural language processing tasks and allows us to assess the performance of privacy-preserving methods on models that process unstructured text data.
\end{itemize}

The target models employed in our experiments are selected to reflect the typical architectures used in their respective domains:

\begin{itemize}
    \item \textbf{AlexNet \cite{krizhevsky2012imagenet}:} Originally developed for the ImageNet Large Scale Visual Recognition Challenge, AlexNet is a deep convolutional neural network consisting of five convolutional layers followed by three fully connected layers. In our experiments on CIFAR-10 and ImageNet-10, AlexNet serves as the target model. Its layered structure, ReLU activations, and use of dropout make it an effective and widely adopted baseline for evaluating both classification performance and membership inference defenses.
    \item \textbf{DistilBERT \cite{sanh2019distilbert}:} It is a compact version of the BERT transformer model, designed to retain much of BERT’s language understanding capabilities while reducing its size and computational cost. In our experiments on SST-2, DistilBERT serves as the target model, offering robust performance on text classification tasks with lower inference latency. Its efficiency makes it well-suited for integrating and testing privacy-preserving mechanisms in the context of Natural Language Processing (NLP).
\end{itemize}

\subsection{Experimental Setup and Results}
We conduct experiments on CIFAR-10, ImageNet-10, and SST-2 datasets. For each dataset, 70\% of the data is used to train and test the target model, while the remaining 30\% is allocated for training shadow and attack models. All models are trained for 30 epochs using stochastic gradient descent (SGD) with a learning rate of 0.01 and a batch size of 64. Experiments are conducted on a machine equipped with NVIDIA RTX 5000 GPU to ensure efficient training and evaluation.

As our experimental baseline defenses, we implement Adversarial Regularization (AdvReg)~\cite{nasr2018machine}, RelaxLoss~\cite{chen2022relaxloss}, MemGuard \cite{jia2019memguard}, SELENA~\cite{tang2022mitigating}, and HAMP~\cite{chen2024overconfidence}. For AdvReg, we adopt the standard min-max training setup with a regularization weight \(\lambda_{\text{reg}} = 3.0\), one adversarial step per batch (\(k_{\text{attack}} = 1\)), and learning rates of \(10^{-2}\) and \(10^{-3}\) for the classifier and attacker, respectively. For RelaxLoss, we follow the loss-smoothing formulation in the original paper and use its key hyperparameters, namely the cross-entropy threshold \(\alpha = 0.5\) and learning rate \(lr = 0.01\). For MemGuard~\cite{jia2019memguard} we follow the original paper’s configuration, using $\varepsilon = 0.5$ and Phase-I coefficients $(c_{2}=10,\ c_{3}=0.1)$, with a defense-classifier trained for $400$ epochs. For SELENA, we use the configuration described in Section~\ref{sec:MIA-Defense}, with \(K=25\) sub-models and \(L=10\) partitions per sample. Finally, for HAMP, we follow Chen et al.~\cite{chen2024overconfidence}, which enforces less-confident
predictions using high-entropy soft labels and an entropy-based regularizer during training, and applies
the testing-time output modification described in the original work. We use the authors' default parameters:
the entropy threshold $\gamma = 0.95$ for generating high-entropy soft labels and the regularization strength
$\alpha = 0.001$ in the training objective.

Each target model is evaluated against multiple types of membership inference attacks, including
Confidence Threshold Attack, Loss Threshold Attack, Entropy-based Attack, M-Entropy Attack, Shadow
Model Attack, and the Likelihood Ratio Attack (LiRA). For each attack, we report the Attack Success
Rate (ASR) under seven conditions: (i) without any defense (None), (ii) AdvReg, (iii) MemGuard,
(iv) RelaxLoss, (v) SELENA, (vi) HAMP, and (vii) our proposed approach DynaNoise.

To quantify the privacy-utility trade-off, we utilize our proposed metric called \emph{MIDPUT}, as
described in Section~\ref{sec:metrics}. MIDPUT measures how effectively a defense reduces membership
inference attack success rates while preserving the predictive accuracy of the target model. We
report both the overall MIDPUT score and per-attack values
(\(\text{MIDPUT}_{Conf}\), \(\text{MIDPUT}_{Loss}\), \(\text{MIDPUT}_{Ent}\),
\(\text{MIDPUT}_{M\text{-}Ent}\), \(\text{MIDPUT}_{Shadow}\), and \(\text{MIDPUT}_{LiRA}\)) for
each evaluated defense.

Tables~\ref{tab:reported_attacks} and~\ref{tab:reported_midput} summarize the experimental results
across all datasets under a diverse set of defense mechanisms. Overall, DynaNoise demonstrates a
consistently strong balance between privacy protection and model utility, achieving substantial
reductions in membership inference attack success rates while preserving near-baseline target model
accuracy across both vision and NLP workloads. In contrast to several prior defenses that obtain
privacy gains primarily through aggressive output suppression or retraining-based regularization,
DynaNoise maintains stable performance across datasets and attack types by adapting the magnitude
of noise to the sensitivity of each individual query.

On CIFAR-10, DynaNoise achieves the highest overall MIDPUT score across all evaluated attacks,
outperforming training-time defenses such as AdvReg and RelaxLoss, post-processing approaches such
as MemGuard, and the ensemble-based SELENA method. Although HAMP reduces all attack success rates to
chance level, it does so by noticeably degrading target model accuracy, resulting in near-zero
net MIDPUT improvement. This outcome highlights a limitation of global confidence suppression on
vision models, where uniformly flattening prediction distributions can unnecessarily distort
low-risk queries and lead to avoidable utility loss.

A similar behavior is observed on ImageNet-10. HAMP again attains low attack success rates under
strong attacks, including Shadow and LiRA, but at the cost of severe accuracy degradation, yielding
the lowest MIDPUT scores among all evaluated defenses. Adversarial Regularization, RelaxLoss, and
MemGuard provide only marginal or inconsistent privacy-utility improvements, while SELENA offers
moderate protection but suffers from reduced accuracy and nontrivial computational overhead. In
contrast, DynaNoise preserves near-baseline accuracy while consistently reducing attack success
rates, resulting in the highest overall MIDPUT score on this dataset and demonstrating robustness
under strong membership inference attacks.

On SST-2, where membership signals are weaker and modern pretrained language
models exhibit relatively high-entropy and well-calibrated predictions, the
behavior of defenses differs from that observed on vision datasets. In this
setting, HAMP attains the highest overall MIDPUT score by reducing attack success
rates to near-chance levels while slightly improving target model accuracy,
which aligns well with its global confidence smoothing strategy in an already
well-calibrated prediction setting. DynaNoise also substantially reduces attack
success rates and preserves near-baseline accuracy, but achieves lower overall
MIDPUT values than HAMP on this dataset. Notably, DynaNoise does so without
uniform confidence suppression, instead selectively injecting noise only for
low-entropy, higher-risk predictions.

\noindent\textbf{A notable pattern is particularly evident on SST-2:}
Several training-based defenses such as Adversarial Regularization and MemGuard provide little to no
privacy-utility improvement, yielding near-zero or negative MIDPUT gains. This behavior is consistent
with the design assumptions discussed in the original AdvReg~\cite{nasr2018machine} and
MemGuard~\cite{jia2019memguard} works, which rely on pronounced separability between member and
non-member predictions to supply a meaningful adversarial signal. For modern pretrained NLP models
such as DistilBERT, member and non-member outputs are already highly similar and well calibrated,
leaving limited room for these defenses to operate effectively. In contrast, defenses that act
directly on prediction calibration, including RelaxLoss, SELENA, HAMP, and DynaNoise, remain
effective in this setting.

\begin{table*}[!htb]
\centering
\caption{Model accuracy and MIA success rates on different datasets under various defense mechanisms.}
\small
\begin{tabular}{|l|l|c|c|c|c|c|c|c|}
\hline
\textbf{Dataset} & \textbf{Defense} & \textbf{Model (\(\uparrow\))} & \textbf{Confidence (\(\downarrow\))} & \textbf{Loss (\(\downarrow\))} & \textbf{Shadow (\(\downarrow\))} & \textbf{LiRA (\(\downarrow\))} & \textbf{Entropy (\(\downarrow\))} & \textbf{M-Entropy (\(\downarrow\))} \\
\hline
\multirow{7}{*}{CIFAR10} 
  & None         & 0.7875 & 0.6124 & 0.6180 & 0.6387 & 0.6097 & 0.6132 & 0.6386 \\
  & AdvReg \cite{nasr2018machine}      & 0.7573 & 0.5516 & 0.5829 & 0.5708 & 0.5836 & 0.5482 & 0.5686 \\
  & MemGuard \cite{jia2019memguard}    & 0.7875 & 0.5948 & 0.6109 & 0.6122 & 0.5893 & 0.5814 & 0.6156 \\
  & RelaxLoss \cite{chen2022relaxloss} & \textbf{0.7890} & 0.5555 & 0.5845 & 0.5665 & 0.5905 & 0.5433 & 0.5639 \\
  & SELENA \cite{tang2022mitigating}   & 0.7674 & 0.5394 & 0.5173 & 0.5346 & 0.5164 & 0.5309 & 0.5326 \\
  & HAMP \cite{chen2024overconfidence} & 0.7422 & \textbf{0.5000} & \textbf{0.5000} & \textbf{0.5000} & \textbf{0.5000} & \textbf{0.5000} & \textbf{0.5000} \\
  & \textbf{DynaNoise}                 & 0.7807 & 0.5014 & 0.5219 & 0.5053 & 0.5342 & 0.5016 & 0.5049 \\
\hline
\multirow{7}{*}{ImageNet-10} 
  & None         & 0.7958 & 0.5824 & 0.6003 & 0.5817 & 0.5721 & 0.5357 & 0.5855 \\
  & AdvReg \cite{nasr2018machine}      & 0.7315 & 0.5244 & 0.5371 & 0.5426 & 0.5381 & 0.5312 & 0.5429 \\
  & MemGuard \cite{jia2019memguard}    & 0.7958 & 0.5800 & 0.6016 & 0.5817 & 0.5721 & 0.5333 & 0.5831 \\
  & RelaxLoss \cite{chen2022relaxloss} & 0.7646 & 0.5203 & 0.5460 & 0.5416 & 0.5436 & 0.5196 & 0.5395 \\
  & SELENA \cite{tang2022mitigating}   & 0.6804 & \textbf{0.4952} & 0.5010 & 0.5055 & 0.5034 & 0.5175 & 0.4993 \\
  & HAMP \cite{chen2024overconfidence} & 0.6588 & 0.5000 & \textbf{0.5000} & \textbf{0.5000} & \textbf{0.5045} & \textbf{0.5000} & \textbf{0.5003} \\
  & \textbf{DynaNoise}                 & \textbf{0.7962} & 0.5316 & 0.5944 & 0.5549 & 0.5598 & 0.5139 & 0.5543 \\
\hline
\multirow{7}{*}{SST-2} 
  & None         & 0.8635 & 0.5162 & 0.5335 & 0.5571 & 0.5364 & 0.5558 & 0.5666 \\
  & AdvReg \cite{nasr2018machine}      & 0.7092 & 0.5126 & 0.5037 & 0.5034 & 0.5154 & \textbf{0.5000} & \textbf{0.5000} \\
  & MemGuard \cite{jia2019memguard}    & 0.8635 & 0.5144 & 0.5335 & 0.5266 & 0.5364 & 0.5252 & 0.5311 \\
  & RelaxLoss \cite{chen2022relaxloss} & 0.8394 & 0.5115 & 0.5125 & \textbf{0.5000} & 0.5147 & \textbf{0.5000} & \textbf{0.5000} \\
  & SELENA \cite{tang2022mitigating}   & \textbf{0.8805} & 0.5270 & 0.5276 & 0.5511 & 0.5234 & 0.5351 & 0.5409 \\
  & HAMP \cite{chen2024overconfidence} & 0.8900 & \textbf{0.5000} & \textbf{0.5000} & \textbf{0.5000} & \textbf{0.5000} & \textbf{0.5000} & \textbf{0.5000} \\
  & \textbf{DynaNoise}                 & 0.8612 & \textbf{0.5000} & 0.5014 & \textbf{0.5000} & 0.5023 & \textbf{0.5000} & \textbf{0.5000} \\
\hline
\end{tabular}
\label{tab:reported_attacks}
\end{table*}

\begin{table*}[!htb]
\centering
\caption{Proposed MIDPUT metrics on different datasets under various defense mechanisms.}
\resizebox{\textwidth}{!}{
\begin{tabular}{|l|l|c|c|c|c|c|c|c|}
\hline
\textbf{Dataset} & \textbf{Defense} & \(\text{MIDPUT}_{Conf} (\uparrow)\) & \(\text{MIDPUT}_{Loss} (\uparrow)\) & \(\text{MIDPUT}_{Shadow} (\uparrow)\) & \(\text{MIDPUT}_{LiRA} (\uparrow)\) & \(\text{MIDPUT}_{Ent} (\uparrow)\) & 
\(\text{MIDPUT}_{M\text{-}Ent} (\uparrow)\) &  \(\text{MIDPUT}_{Overall} (\uparrow)\) \\
\hline
\multirow{6}{*}{CIFAR10} 
  & AdvReg \cite{nasr2018machine}      & 0.0306 & 0.0049 & 0.0377 & -0.0041 & 0.0348 & 0.0399 & 0.0240 \\
  & MemGuard \cite{jia2019memguard}    & 0.0176 & 0.0071 & 0.0086 & -0.0001 & 0.0125 & 0.0110 & 0.0095 \\
  & RelaxLoss \cite{chen2022relaxloss} & 0.0584 & 0.0350 & 0.0737 & 0.0207 & 0.0714 & 0.0762 & 0.0559 \\
  & SELENA \cite{tang2022mitigating}   & 0.0529 & 0.0806 & 0.0840 & \textbf{0.0732} & 0.0623 & 0.0860 & 0.0732 \\
  & HAMP \cite{chen2024overconfidence} & 0.0571 & 0.0627 & 0.0834 & 0.0544 & 0.0579 & 0.0833 & 0.0665 \\
  & \textbf{DynaNoise}                 & \textbf{0.1043} & \textbf{0.0893} & \textbf{0.1266} & 0.0687 & \textbf{0.1048} & \textbf{0.1270} & \textbf{0.1035} \\
\hline
\multirow{6}{*}{ImageNet-10} 
  & AdvReg \cite{nasr2018machine}      & -0.0062 & -0.0010 & -0.0251 & -0.0302 & -0.0598 & -0.0216 & -0.0240 \\
  & MemGuard \cite{jia2019memguard}    & 0.0024 & -0.0014 & 0.0000 & 0.0000 & 0.0024 & 0.0024 & 0.0010 \\
  & RelaxLoss \cite{chen2022relaxloss} & 0.0310 & \textbf{0.0231} & 0.0090 & -0.0027 & -0.0150 & 0.0149 & 0.0101 \\
  & SELENA \cite{tang2022mitigating}   & -0.0282 & -0.0161 & -0.0391 & -0.0467 & -0.0972 & -0.0292 & -0.0428 \\
  & HAMP \cite{chen2024overconfidence} & -0.0545 & -0.0233 & -0.0463 & -0.0624 & -0.0806 & -0.0397 & -0.0511 \\
  & \textbf{DynaNoise}                 & \textbf{0.0512} & 0.0062 & \textbf{0.0272} & \textbf{0.0127} & \textbf{0.0222} & \textbf{0.0316} & \textbf{0.0252} \\
\hline
\multirow{6}{*}{SST-2} 
  & AdvReg \cite{nasr2018machine}      & -0.1507 & -0.1245 & -0.1006 & -0.1333 & -0.0985 & -0.0877 & -0.1159 \\
  & MemGuard \cite{jia2019memguard}    & 0.0019 & 0.0000 & 0.0305 & 0.0000 & 0.0306 & 0.0354 & 0.0164 \\
  & RelaxLoss \cite{chen2022relaxloss} & -0.0193 & -0.0031 & 0.0332 & -0.0023 & 0.0317 & 0.0425 & 0.0138 \\
  & SELENA \cite{tang2022mitigating}   & 0.0097 & 0.0264 & 0.0265 & 0.0335 & 0.0412 & 0.0462 & 0.0306 \\
  & HAMP \cite{chen2024overconfidence} & \textbf{0.0427} & \textbf{0.06} & \textbf{0.0836} & \textbf{0.0629} & \textbf{0.0823} & \textbf{0.0931} & \textbf{0.0708} \\

  & \textbf{DynaNoise} & 0.0139 & 0.0298 & 0.0548 & 0.0318 & 0.0535 & 0.0643 & 0.0414 \\
\hline
\end{tabular}}
\label{tab:reported_midput}
\end{table*}

\begin{figure*}[t]
\centering
\begin{subfigure}{0.245\textwidth}
  \centering
  \includegraphics[width=\linewidth]{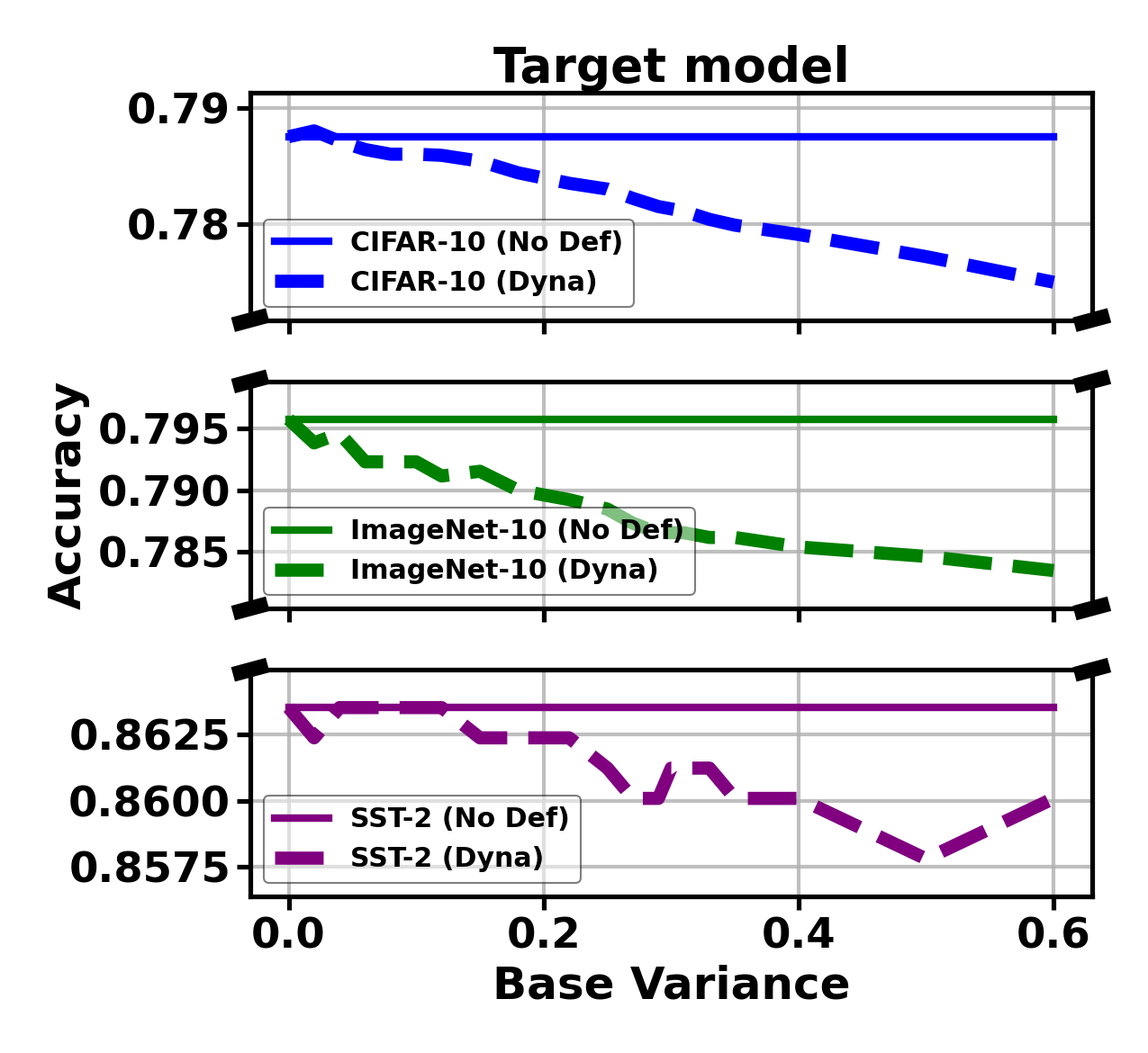}
  \caption{Target model}
\end{subfigure}
\begin{subfigure}{0.245\textwidth}
  \centering
  \includegraphics[width=\linewidth]{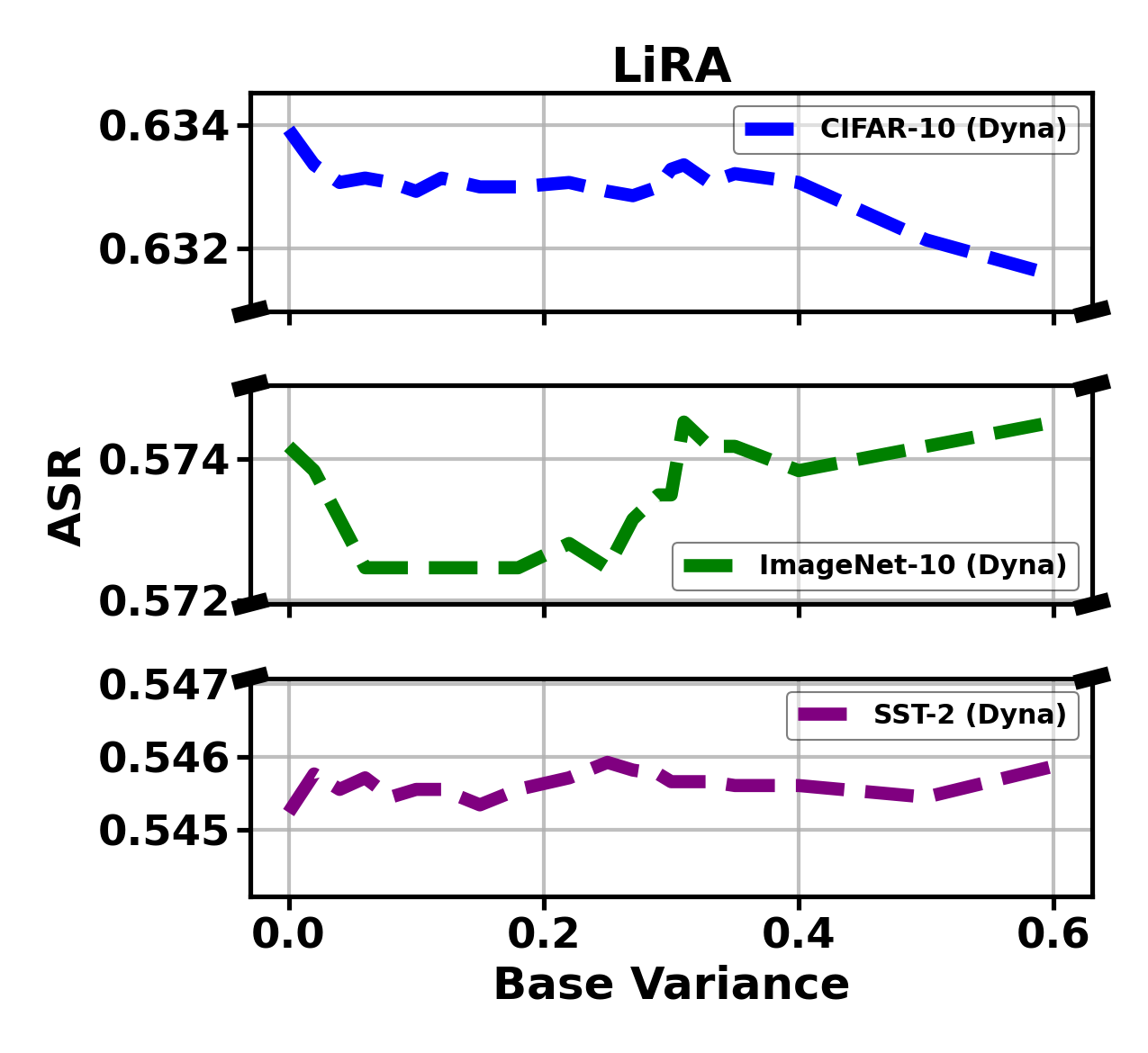}
  \caption{LiRA}
\end{subfigure}
\begin{subfigure}{0.245\textwidth}
  \centering
  \includegraphics[width=\linewidth]{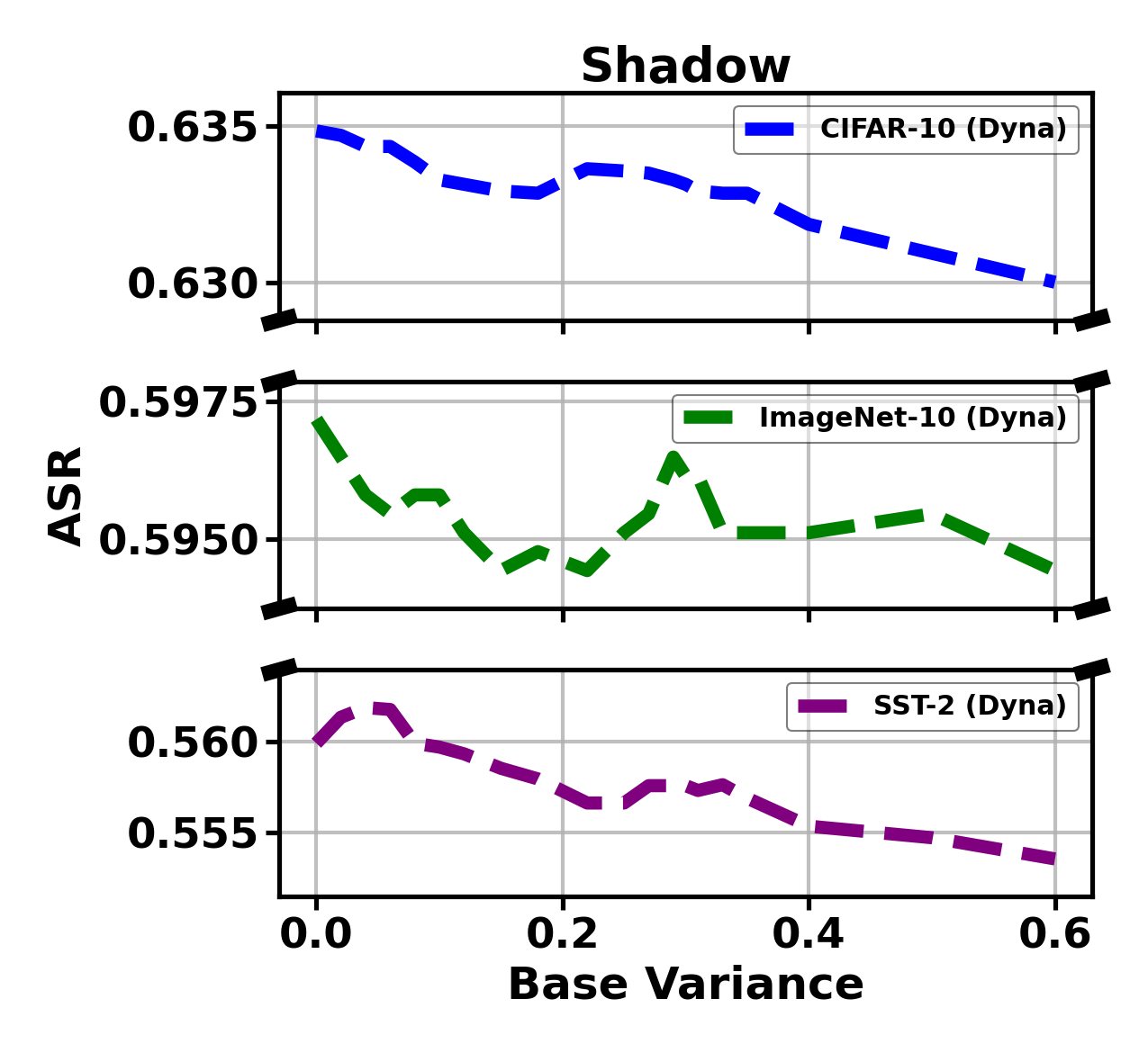}
  \caption{Shadow}
\end{subfigure}
\begin{subfigure}{0.245\textwidth}
  \centering
  \includegraphics[width=\linewidth]{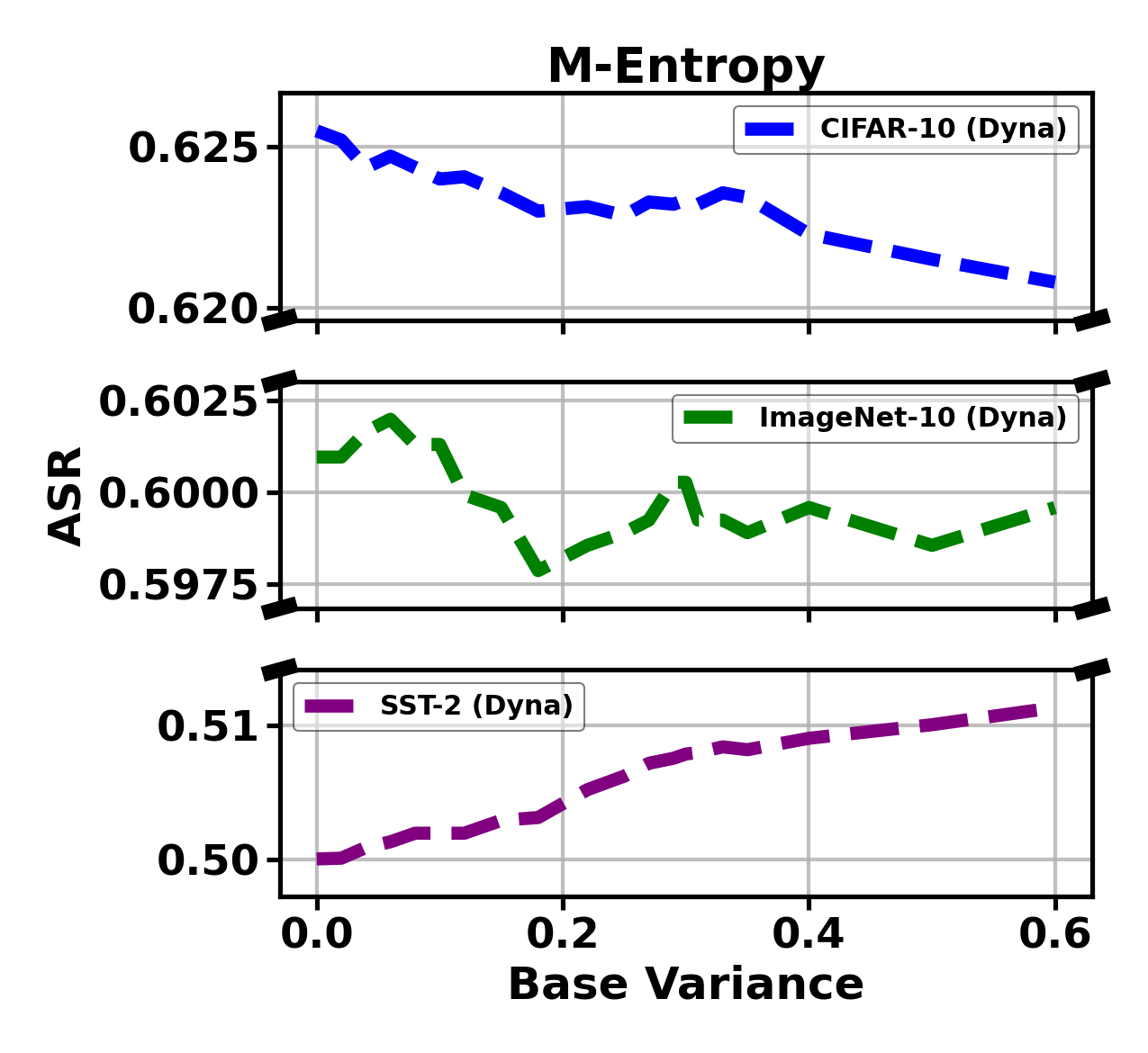}
  \caption{M-Entropy}
\end{subfigure}
\vspace{-2mm}
\caption{\textbf{Base variance sweep.} Target model accuracy and representative attack ASR (before vs.\ after DynaNoise) across CIFAR-10, ImageNet-10, and SST-2.}
\label{fig:bv_main}
\vspace{-2mm}
\end{figure*}

\begin{figure*}[t]
\centering
\begin{subfigure}{0.245\textwidth}
  \centering
  \includegraphics[width=\linewidth]{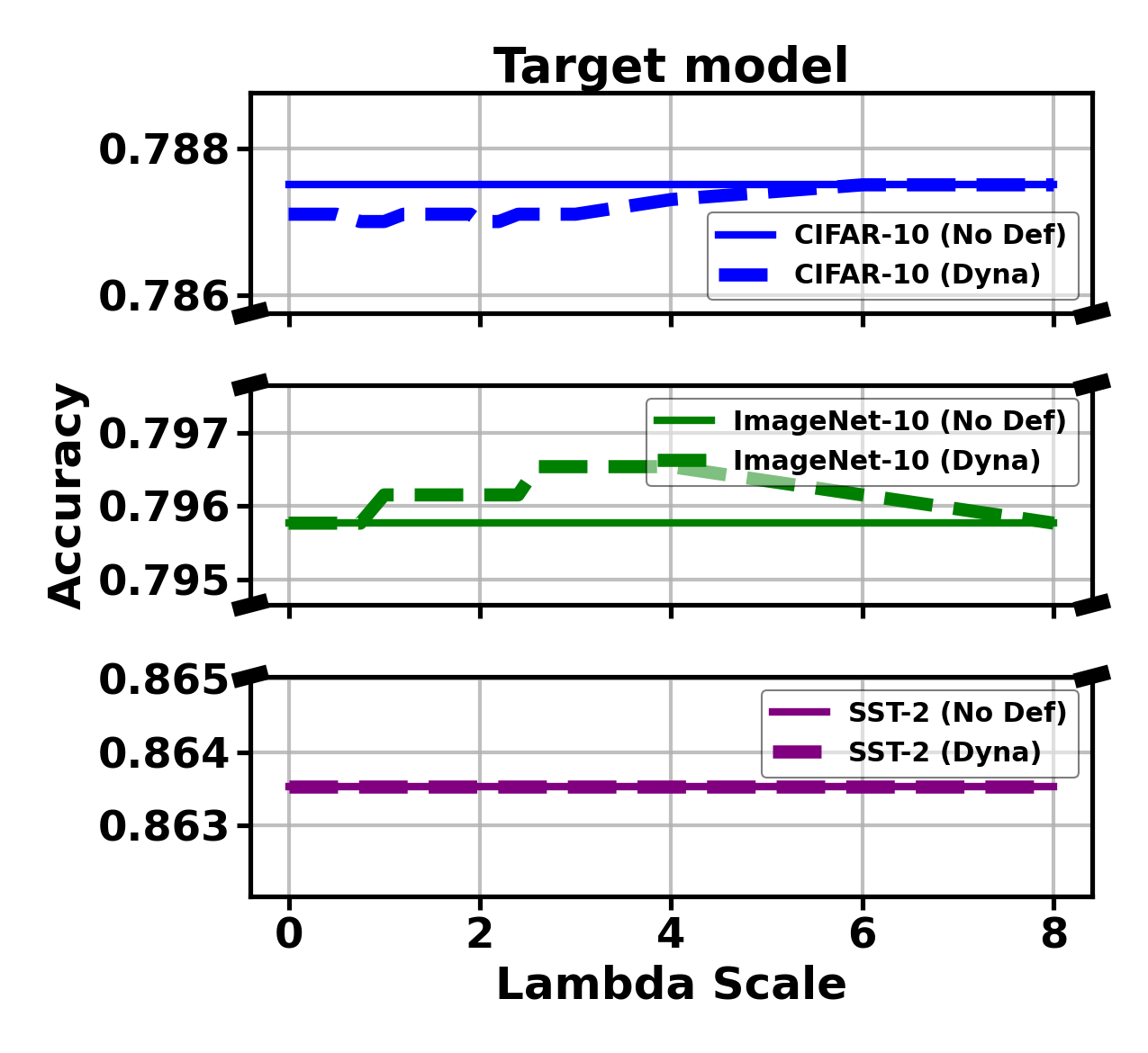}
  \caption{Target model}
\end{subfigure}
\begin{subfigure}{0.245\textwidth}
  \centering
  \includegraphics[width=\linewidth]{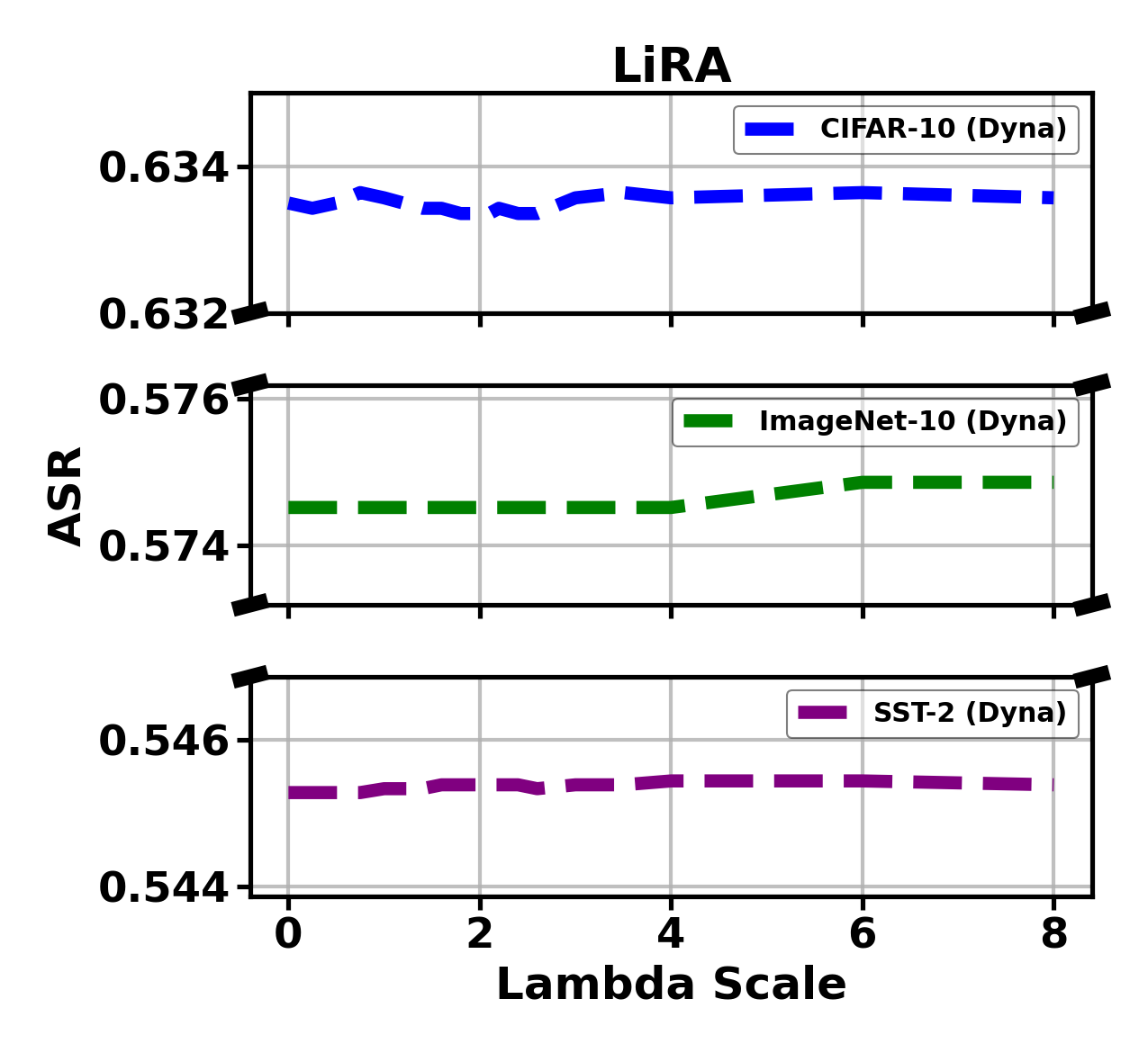}
  \caption{LiRA}
\end{subfigure}
\begin{subfigure}{0.245\textwidth}
  \centering
  \includegraphics[width=\linewidth]{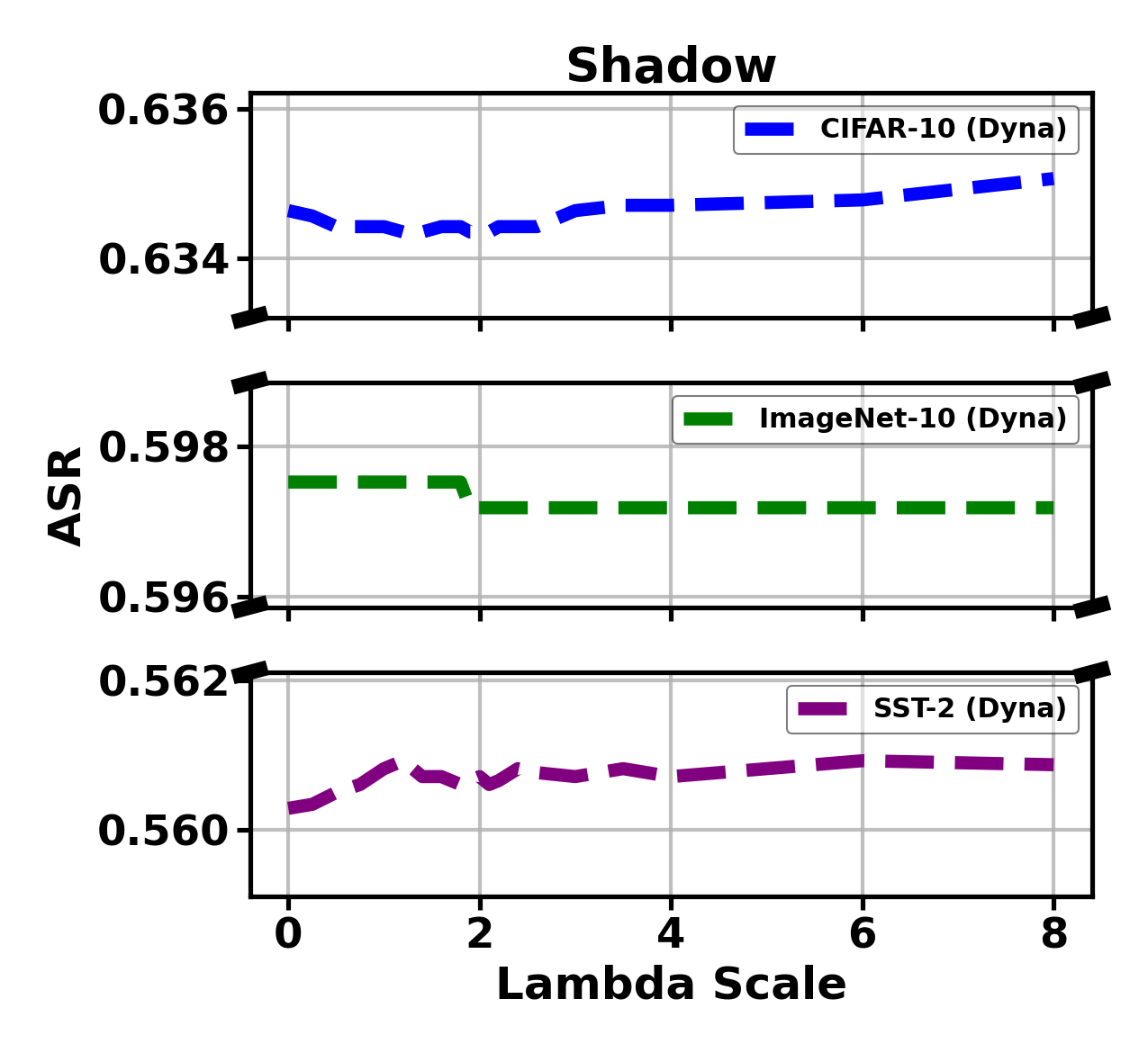}
  \caption{Shadow}
\end{subfigure}
\begin{subfigure}{0.245\textwidth}
  \centering
  \includegraphics[width=\linewidth]{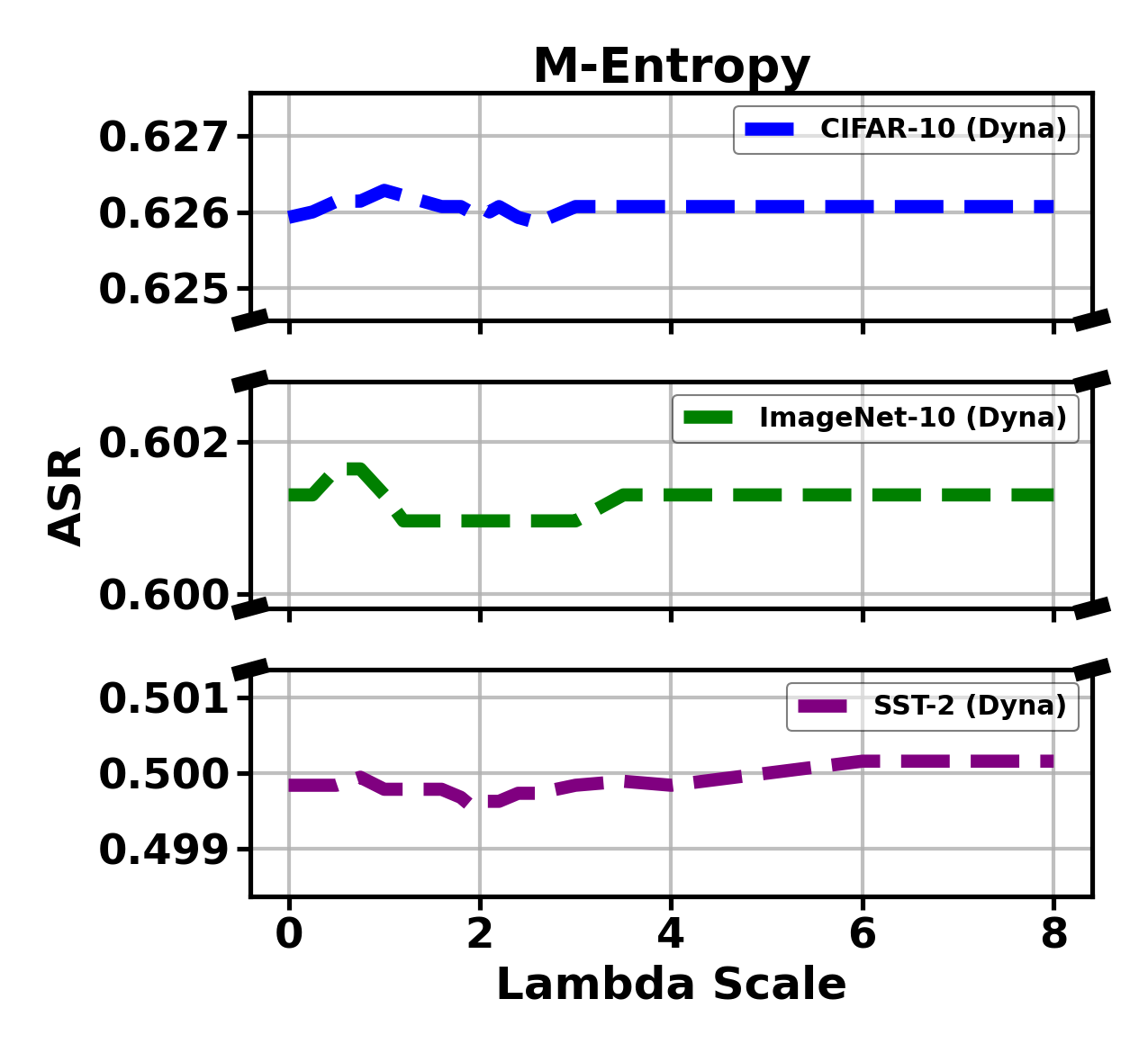}
  \caption{M-Entropy}
\end{subfigure}
\vspace{-2mm}
\caption{\textbf{Lambda scale sweep.} Target model accuracy and representative attack ASR (before vs.\ after DynaNoise) across datasets.}
\label{fig:ls_main}
\vspace{-2mm}
\end{figure*}

\begin{figure*}[t]
\centering
\begin{subfigure}{0.245\textwidth}
  \centering
  \includegraphics[width=\linewidth]{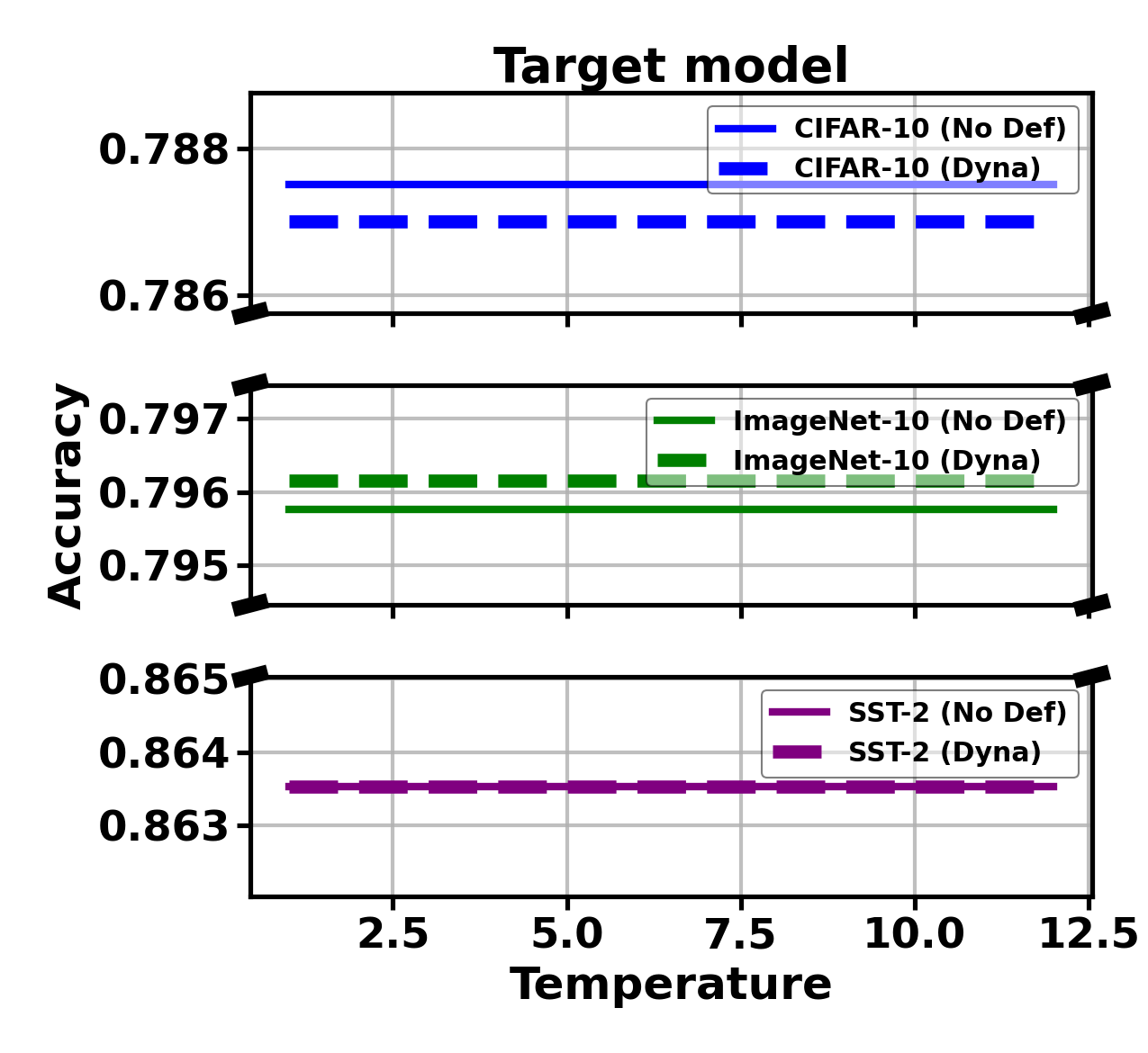}
  \caption{Target model}
\end{subfigure}
\begin{subfigure}{0.245\textwidth}
  \centering
  \includegraphics[width=\linewidth]{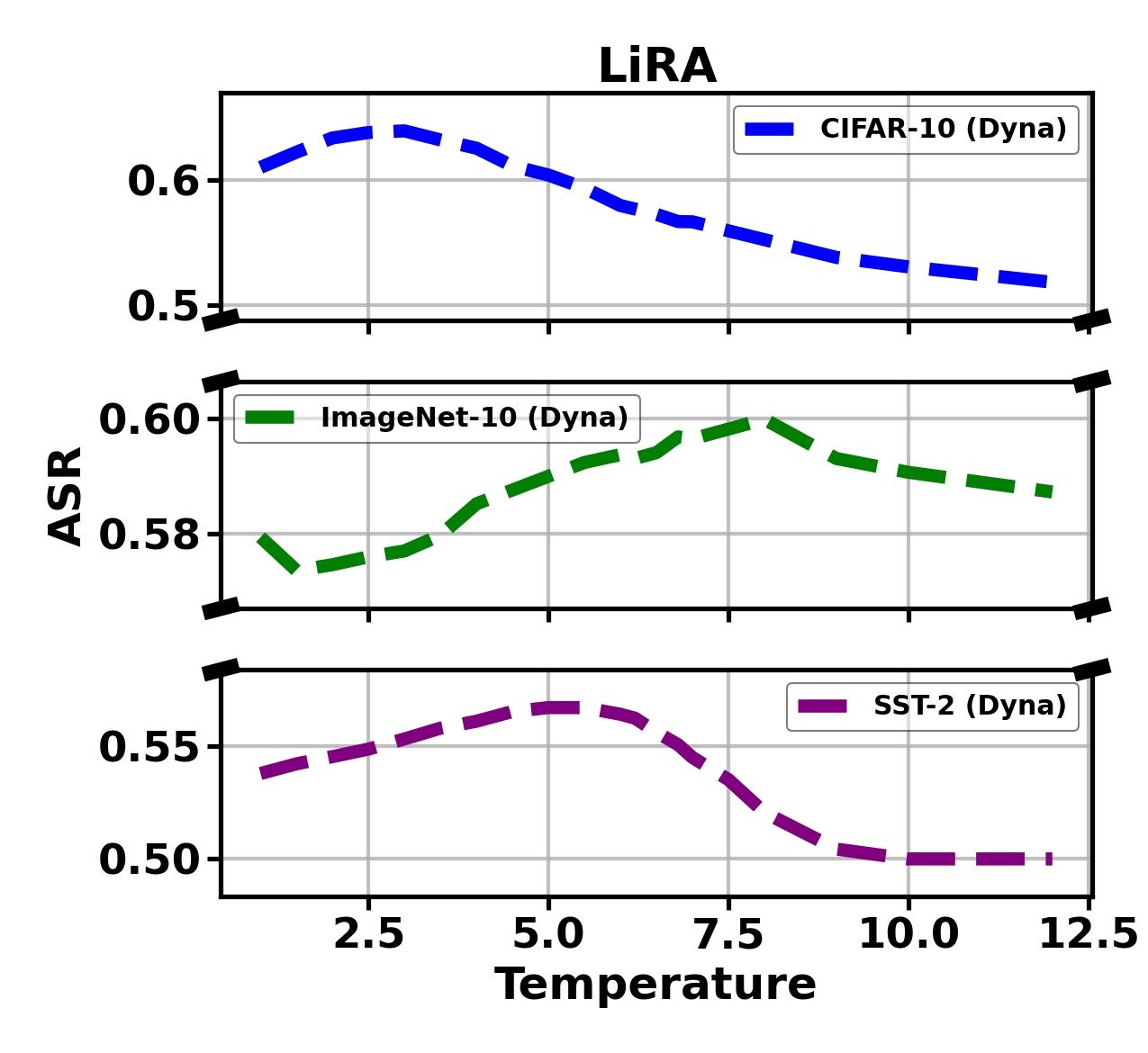}
  \caption{LiRA}
\end{subfigure}
\begin{subfigure}{0.245\textwidth}
  \centering
  \includegraphics[width=\linewidth]{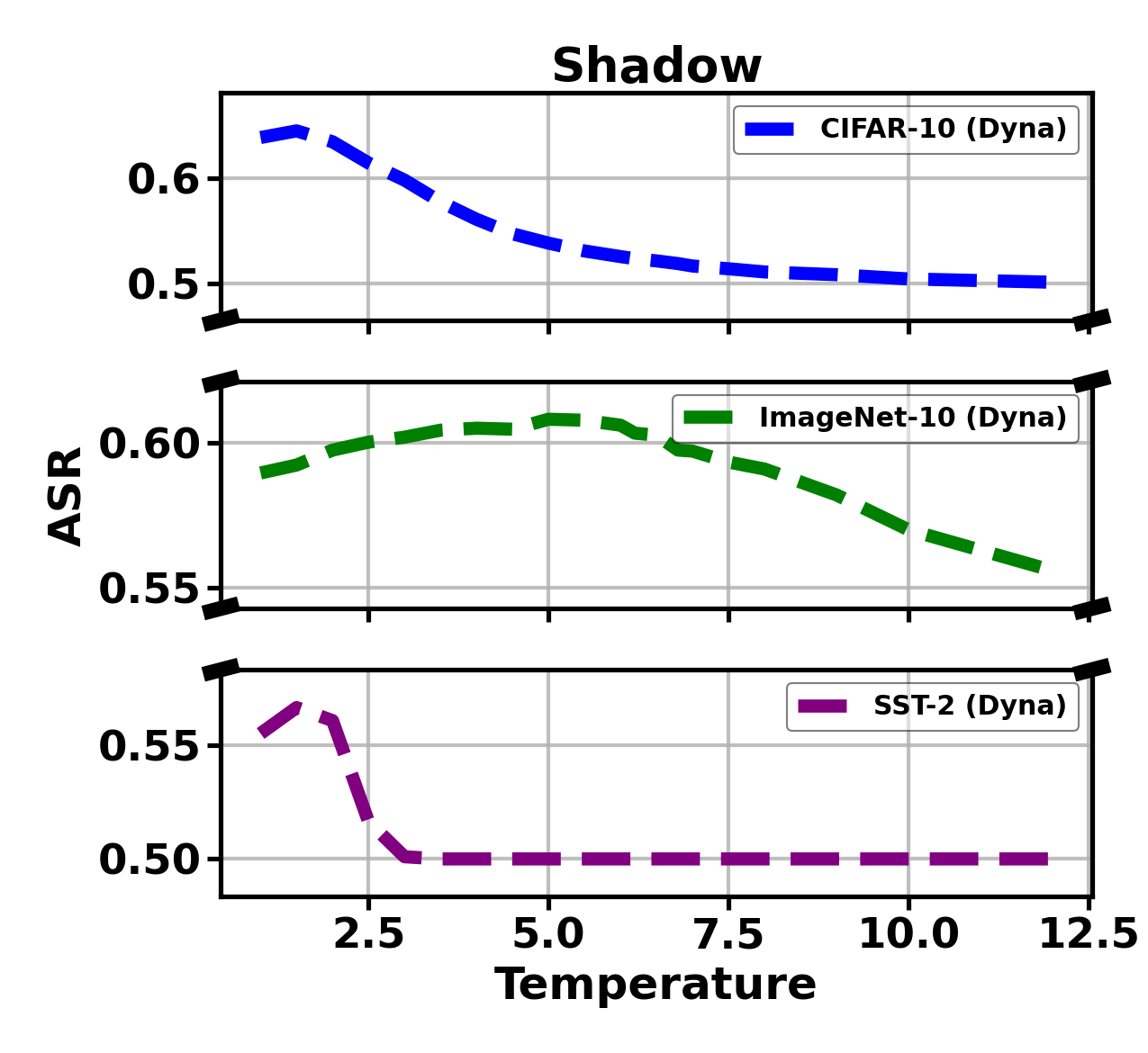}
  \caption{Shadow}
\end{subfigure}
\begin{subfigure}{0.245\textwidth}
  \centering
  \includegraphics[width=\linewidth]{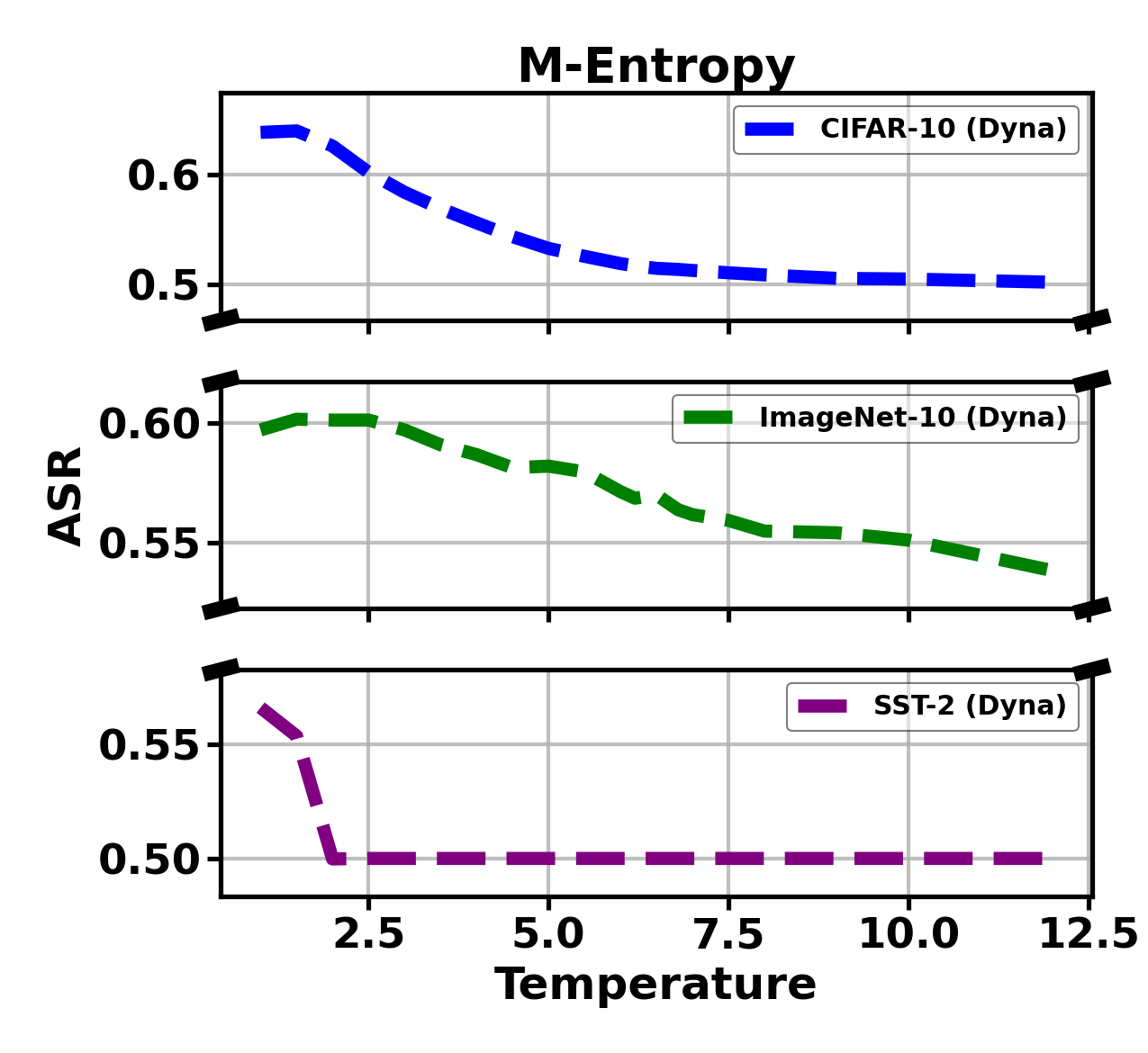}
  \caption{M-Entropy}
\end{subfigure}
\vspace{-2mm}
\caption{\textbf{Temperature sweep.} Target model accuracy and representative attack ASR (before vs.\ after DynaNoise) across datasets.}
\label{fig:t_main}
\vspace{-2mm}
\end{figure*}

Figures~\ref{fig:bv_main}, \ref{fig:ls_main}, and \ref{fig:t_main} show how the three DynaNoise
parameters, base variance, lambda scale, and temperature, affect target model accuracy and
representative membership inference attacks across CIFAR-10, ImageNet-10, and SST-2. Each figure
reports results before and after applying DynaNoise, enabling direct assessment of privacy--utility
trade-offs under parameter variation. Varying the base variance (Figure~\ref{fig:bv_main}) results in only minor changes in target model
accuracy while yielding small but consistent reductions in attack success rates, particularly for
LiRA and Shadow. The M-Entropy attack remains largely stable across this range, suggesting that base
variance influences the overall scale of perturbation while preserving the underlying accuracy
profile.

A similar trend is observed for the lambda scale (Figure~\ref{fig:ls_main}), where increasing the
parameter results in minimal changes to target model accuracy and modest, incremental reductions
in attack success rates. This behavior indicates that the lambda scale primarily adjusts the
magnitude of entropy-guided noise without substantially altering the overall privacy--utility
profile. Temperature variation (Figure~\ref{fig:t_main}) exhibits a more pronounced effect, with
higher temperatures consistently reducing attack success rates under LiRA, Shadow, and M-Entropy
across all evaluated datasets while leaving accuracy largely unchanged. Taken together, these
results suggest that lambda scaling and temperature smoothing influence privacy through different
but complementary mechanisms, jointly shaping the privacy--utility behavior of DynaNoise.

\subsection{Time and Computational Overhead Analysis}

The evaluated defense mechanisms differ substantially in their computational
requirements during both training and inference. Empirical measurements of
training and inference overhead are summarized in Appendix~\ref{app:overhead}.
We next discuss the computational complexity of each defense.

\textbf{DynaNoise} perturbs model logits by adding Gaussian noise whose variance adapts to query sensitivity, followed by a softmax recomputation of the $\ell$-dimensional probability vector. This operation has per-sample cost $\mathcal{O}(\ell)$, corresponding to one noise draw and a single softmax evaluation. As a post-hoc defense, DynaNoise imposes minimal runtime overhead and requires no retraining, making it lightweight and easily deployable.

\textbf{Adversarial Regularization} introduces a min--max training framework in
which an auxiliary membership attack model is jointly optimized with the target
classifier. Each training step includes additional attacker updates, yielding an
overall training complexity of $\mathcal{O}(C_{\text{model}} + k \cdot
C_{\text{attack}})$, where $C_{\text{attack}}$ denotes the attack model update
cost and $k$ the number of attack iterations per batch. As a result, this defense
requires retraining the target model and increases the complexity of the training
procedure, while inference cost remains identical to that of the base
classifier.

\textbf{RelaxLoss} modifies the training objective by penalizing overly confident predictions via a confidence-regularization term. This defense requires no auxiliary models and introduces no stochastic components, resulting in training complexity $\mathcal{O}(C_{\text{model}})$ and unchanged inference cost. Thus, RelaxLoss retains training efficiency comparable to the corresponding undefended baseline.

\textbf{MemGuard} protects against score-based membership inference attacks by solving a constrained optimization problem at inference time to minimally perturb the model’s posterior distribution. For each query, MemGuard performs an inner-loop iterative optimization over the output logits, with complexity approximately $\mathcal{O}(T \cdot \ell)$, where $T$ is the number of gradient steps and $\ell$ is the number of classes. While model retraining is not required, inference-time overhead is significantly higher than DynaNoise, RelaxLoss, or AdvReg, since a full iterative procedure must be run for every input. This makes MemGuard substantially more computationally expensive and less suited for real-time or large-scale deployments.

\textbf{SELENA} trains an ensemble of $K$ sub-models under the Split-AI framework and performs self-distillation to aggregate their predictions. This results in a total training complexity of $\mathcal{O}(K \cdot C_{\text{model}})$ and a correspondingly large memory footprint. After distillation, inference cost becomes equivalent to a single model, but the training phase is considerably more resource-intensive. Consistent with this design, empirical inference latency after distillation
matches the baseline model (Appendix~\ref{app:overhead}).

\textbf{HAMP} combines training-time entropy regularization with inference-time output modification.
During training, HAMP enforces high-entropy soft labels and an entropy-maximization term in the loss,
resulting in training complexity $\mathcal{O}(C_{\text{model}})$ with no auxiliary models. At
inference time, HAMP generates random reference inputs and replaces the sorted output probabilities
of the queried sample with those from the random inputs while preserving label ordering. This
procedure introduces an additional forward pass per query and a sorting operation over the output
distribution, yielding an inference-time complexity of approximately $\mathcal{O}(C_{\text{model}}
+ \ell \log \ell)$. While HAMP avoids retraining multiple models, its inference overhead is higher
than that of DynaNoise due to the extra forward evaluation and probability
replacement step, yielding approximately a $2\times$ inference-time overhead in
practice (Appendix~\ref{app:overhead}).

\noindent\textbf{Overall.} DynaNoise offers the most computationally efficient defense among all
evaluated methods. Unlike RelaxLoss, AdvReg, SELENA, and HAMP, it operates entirely post-hoc, requires
no retraining, introduces no auxiliary models, and adds only an $\mathcal{O}(\ell)$ inference-time
operation. Compared to MemGuard and HAMP, which incur additional per-query optimization or extra
forward passes, DynaNoise remains substantially lighter at inference time. As a result, DynaNoise
achieves strong privacy protection with the smallest computational and memory footprint, making it
well suited for deployment on pretrained models in resource-constrained or large-scale settings.

\subsection{Discussion and Limitations}
\label{sec:limitations}

DynaNoise provides a lightweight, post-hoc defense against membership
inference attacks, achieving strong privacy protection with minimal impact on
model accuracy and inference cost. We next discuss its practical advantages and
limitations.

\noindent\textbf{Advantages:}
\begin{itemize}
    \item \textbf{Distribution-aware sensitivity:}  
        Unlike SELENA, AdvReg, RelaxLoss, MemGuard, and HAMP, which apply fixed training-time regularization,ensemble aggregation, or global inference-time output modification, DynaNoise adapts its perturbation level using a continuous, entropy-based measure of prediction sensitivity. By leveraging the full predictive distribution on a per-query basis, DynaNoise selectively injects noise for high-risk predictions while avoiding unnecessary perturbations on low-risk queries.

    \item \textbf{Post-hoc and model-agnostic:}  
    The defense operates purely at inference time and requires no retraining, architectural changes, or
    access to training data. In contrast, AdvReg, RelaxLoss, and HAMP modify the training procedure,
    SELENA trains an ensemble of sub-models, and MemGuard solves a per-query constrained optimization
    problem over logits. DynaNoise can therefore be deployed immediately on pretrained models such as
    AlexNet or DistilBERT, making it well suited for MLaaS and black-box settings.

    \item \textbf{Efficiency and simplicity:}  
    DynaNoise introduces only three interpretable hyperparameters (\(\sigma_0\), \(\lambda\), and \(T\))
    and adds a single noise sampling and softmax evaluation per query, resulting in
    $\mathcal{O}(\ell)$ inference-time overhead. In comparison, RelaxLoss and HAMP require retraining
    with modified objectives, SELENA incurs ensemble training costs, AdvReg relies on adversarial
    optimization, and MemGuard performs iterative per-query perturbation solving. As a result,
    DynaNoise remains the lightest-weight option, with the lowest computational and memory footprint
    among all evaluated defenses.

    \item \textbf{Superior privacy-utility trade-off:}  
    Across CIFAR-10 and ImageNet-10, DynaNoise achieves the highest overall MIDPUT values, delivering
    strong reductions in attack success rates while preserving near-baseline target model accuracy.
    On SST-2, DynaNoise provides competitive privacy gains with stable accuracy, closely matching the
    best-performing defenses. Compared to SELENA, AdvReg, RelaxLoss, MemGuard, and HAMP, DynaNoise
    consistently exhibits favorable privacy-utility behavior across diverse datasets and attack types.

\end{itemize}

\noindent\textbf{Limitations:}
\begin{itemize}
    
    \item \textbf{Hyperparameter tuning:}  
    Similar to SELENA, AdvReg, RelaxLoss, MemGuard, and HAMP, the effectiveness of DynaNoise depends on
    selecting appropriate hyperparameters, including its variance-scaling and temperature components.
    While we provide stable default ranges that perform well across datasets, some dataset-specific
    tuning may be required to achieve optimal privacy-utility trade-offs.

    \item \textbf{No formal privacy guarantees:}  
    Like SELENA, AdvReg, RelaxLoss, MemGuard, and HAMP, DynaNoise provides empirical robustness against
    membership inference attacks but does not offer formal differential privacy guarantees. Developing
    theoretical privacy bounds for entropy-aware, query-adaptive output perturbation remains an
    important direction for future work.

    \item \textbf{Vulnerability to repeated queries:}
    As with other stochastic inference-time defenses, including MemGuard, HAMP, and
    DynaNoise, repeated or highly similar queries may allow an adversary to average
    out injected randomness and partially recover the underlying output distribution
    \cite{rahimian2020sampling}. DynaNoise therefore assumes a limited-query threat
    model; potential mitigations include enforcing per-record query limits and
    detecting suspicious query patterns using stateful or similarity-based monitors
    \cite{li2022blacklight,li2024qpa}.

   \item \textbf{Label-only membership inference:}
DynaNoise protects against score-based membership inference attacks and does not
defend against label-only MIAs that rely solely on predicted labels. Extending
defenses to this threat model is an important direction for future work.

\end{itemize}

\section{Conclusion}
\label{sec:conclusion}
In this work, we introduced DynaNoise, a lightweight and adaptive inference-time defense against membership inference attacks. Unlike prior approaches that require retraining, ensemble models, or uniform confidence suppression, DynaNoise operates as a purely post-hoc mechanism that dynamically injects calibrated noise based on query sensitivity. By modulating protection strength using entropy, DynaNoise selectively shields high-risk predictions while preserving the utility of low-risk queries. Extensive evaluation across vision and language benchmarks demonstrates that this query-adaptive strategy consistently reduces attack success rates while maintaining near-baseline accuracy, outperforming training-time, post-processing, and ensemble-based defenses in terms of overall privacy–utility trade-off.

To facilitate holistic evaluation of inference-time defenses, we also proposed the MIDPUT metric as a complementary summary measure capturing the balance between privacy protection and predictive performance. Across diverse datasets and attack settings, DynaNoise achieves strong MIDPUT values, highlighting the effectiveness of adapting protection to query sensitivity rather than applying global perturbations. These results indicate that query-aware noise injection offers a practical and resource-efficient path toward improving privacy in deployed machine learning systems. Future work will explore extending the DynaNoise framework with alternative sensitivity estimators, such as modified entropy, and integrating adaptive noise mechanisms with formal privacy techniques to further strengthen robustness while retaining strong empirical performance.

\bibliographystyle{ACM-Reference-Format}
\bibliography{sample-base}

\appendix
\section{Entropy--Sensitivity Correlation Plots}
\label{app:entropy-sensitivity-plots}

\begin{figure*}[t]
    \centering
    \begin{subfigure}[b]{0.32\linewidth}
        \centering
        \includegraphics[width=\linewidth]{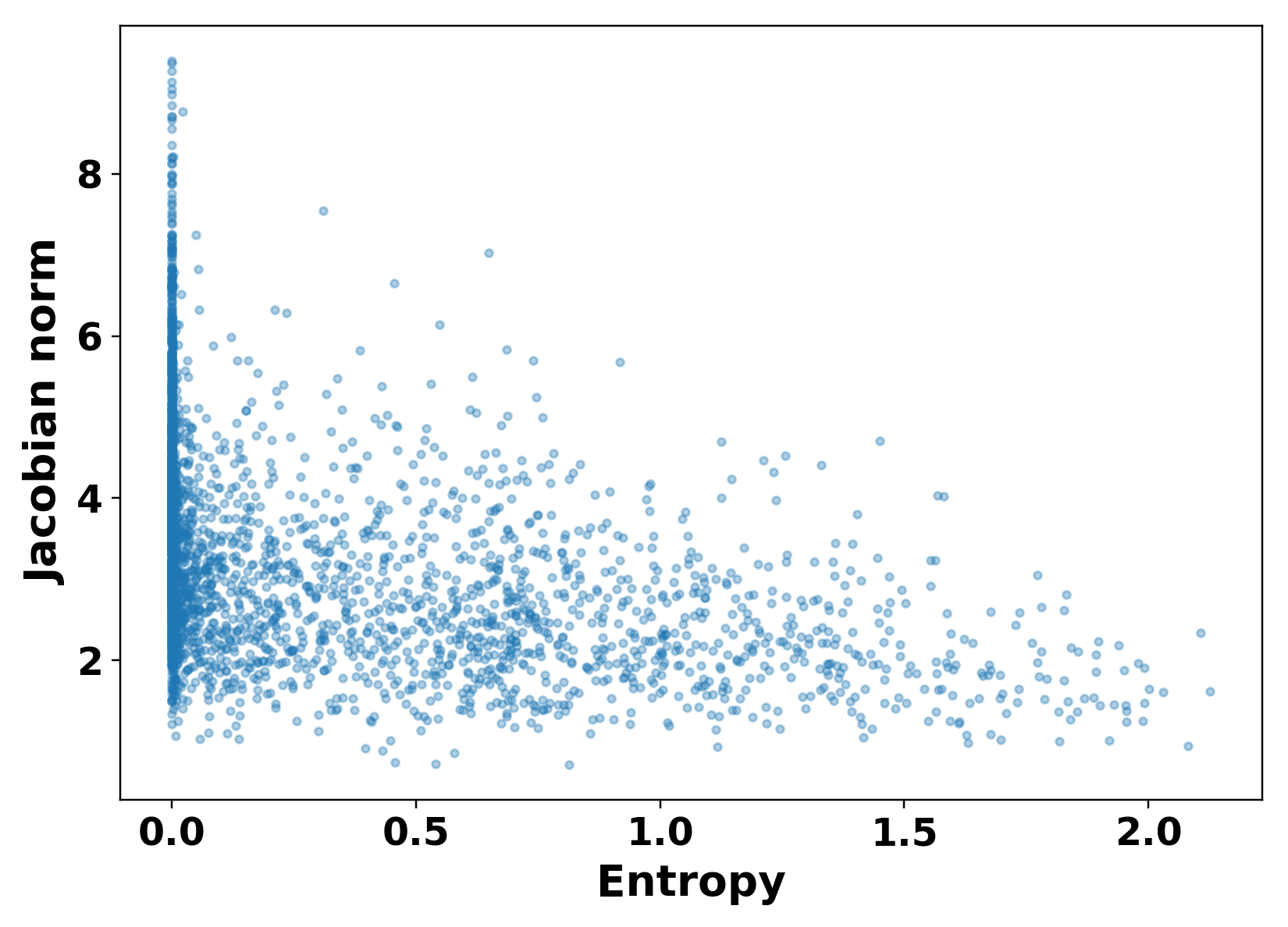}
        \caption{CIFAR-10 (AlexNet), $r=-0.41$}
    \end{subfigure}
    \hfill
    \begin{subfigure}[b]{0.32\linewidth}
        \centering
        \includegraphics[width=\linewidth]{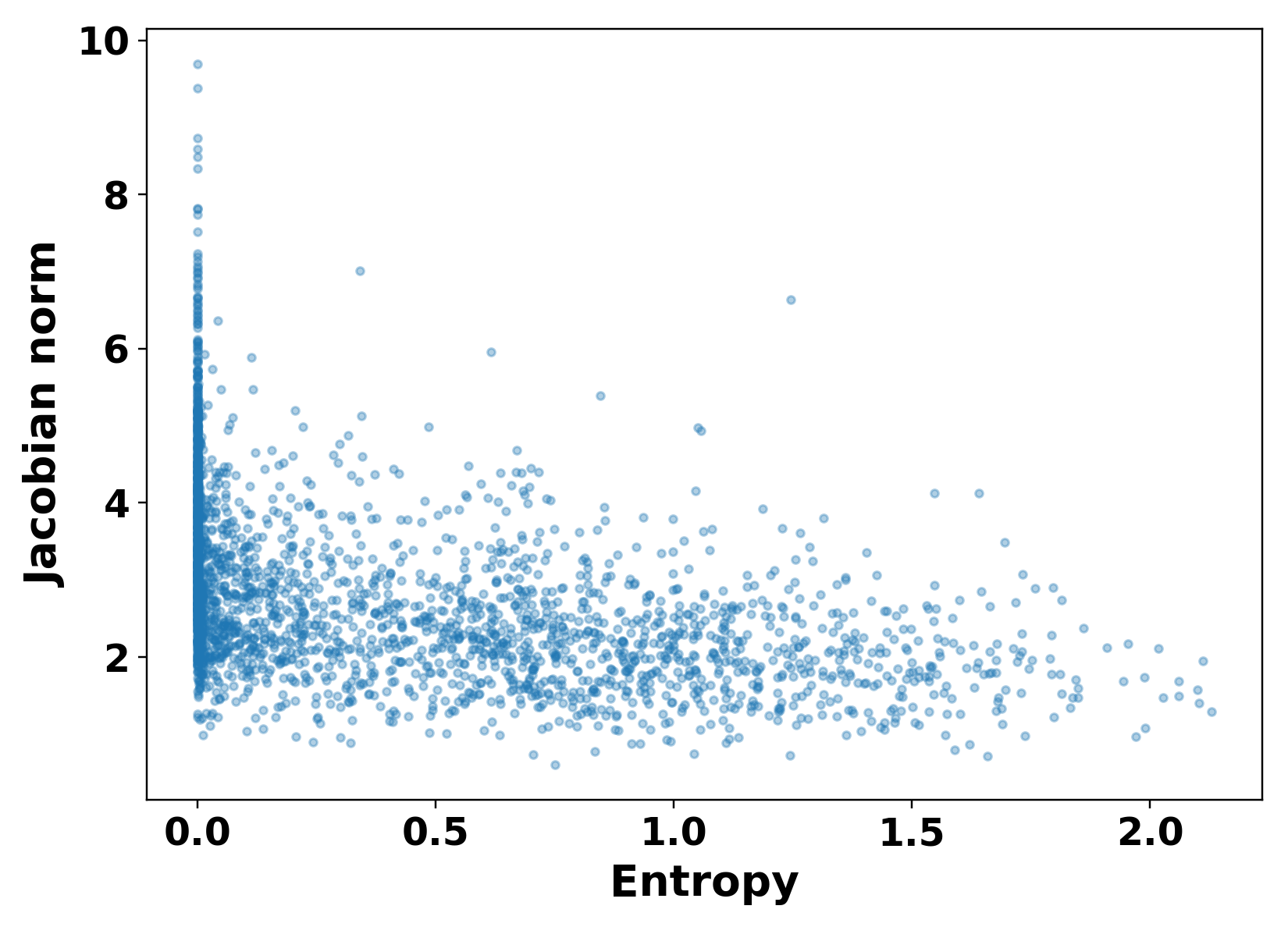}
        \caption{ImageNet-10 (AlexNet), $r=-0.46$}
    \end{subfigure}
    \hfill
    \begin{subfigure}[b]{0.32\linewidth}
        \centering
        \includegraphics[width=\linewidth]{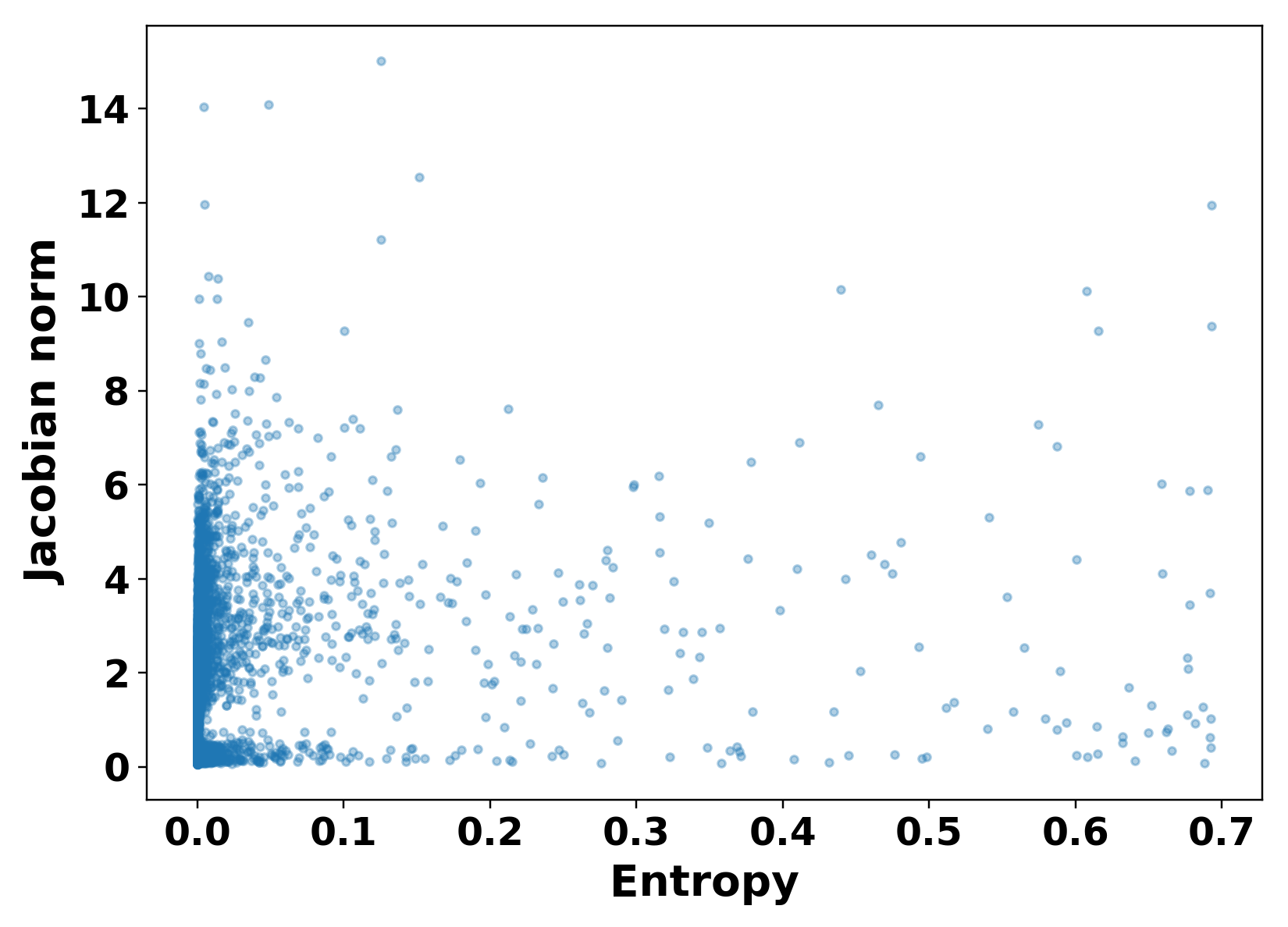}
        \caption{SST-2 (DistilBERT), $r=0.196$}
    \end{subfigure}

    \caption{
        Correlation between output entropy and Jacobian-based sensitivity across datasets and model architectures. These plots support the analysis in Section~\ref{sec:entropy-jacobian-connection}.
    }
    \label{fig:entropy_sensitivity_stack}
\end{figure*}

\noindent
This section contains the full entropy--Jacobian sensitivity correlation plots, which provide empirical support for the analysis in Section~\ref{sec:entropy-jacobian-connection}, as illustrated in Figure~\ref{fig:entropy_sensitivity_stack}.

\section{Dataset-Level Confidence and Entropy Analysis}
\label{app:confidence-entropy}

\begin{figure*}[t]
    \centering
    \begin{minipage}{0.48\linewidth}
        \centering
        \includegraphics[width=0.48\linewidth]{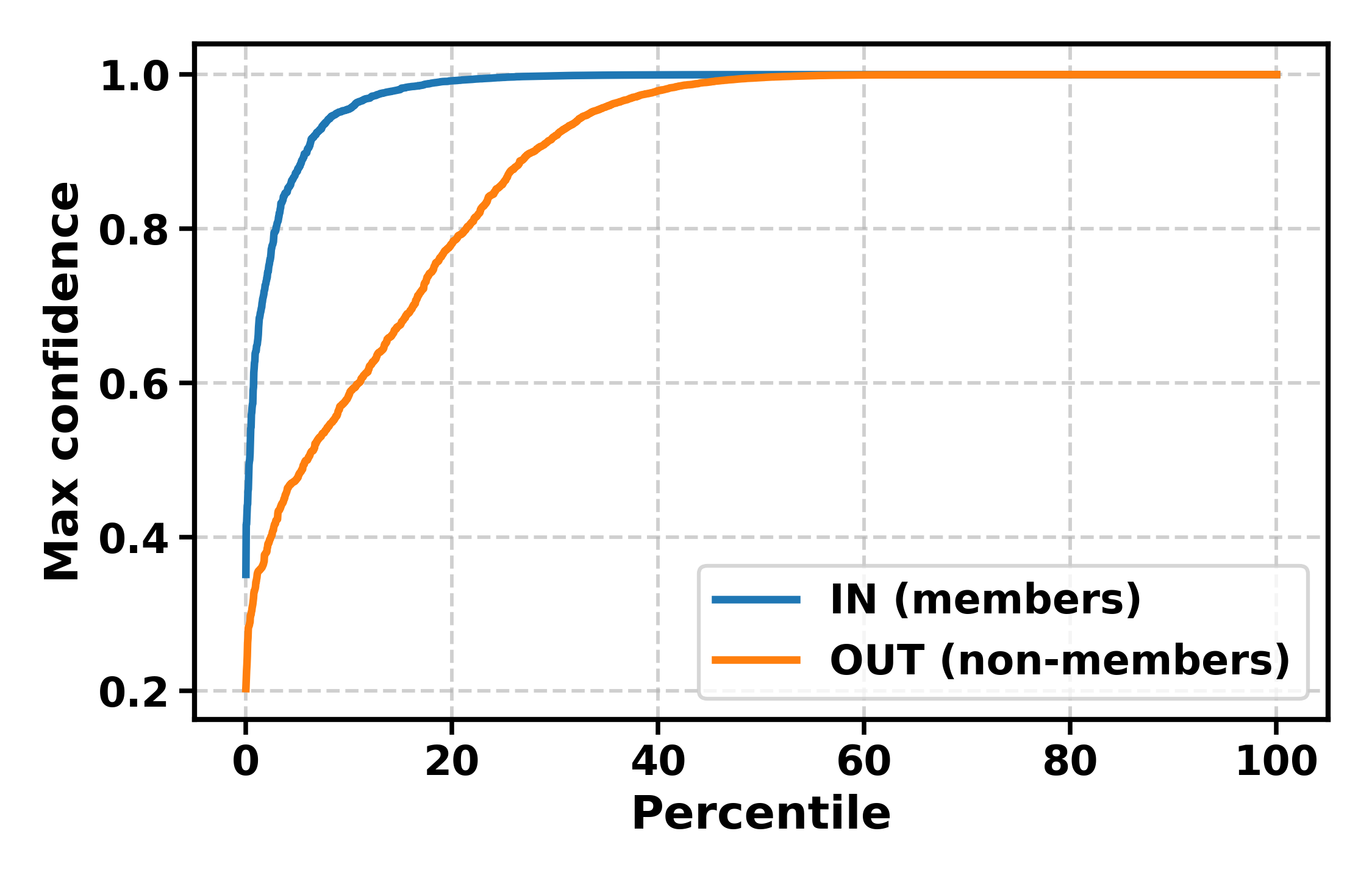}
        \includegraphics[width=0.48\linewidth]{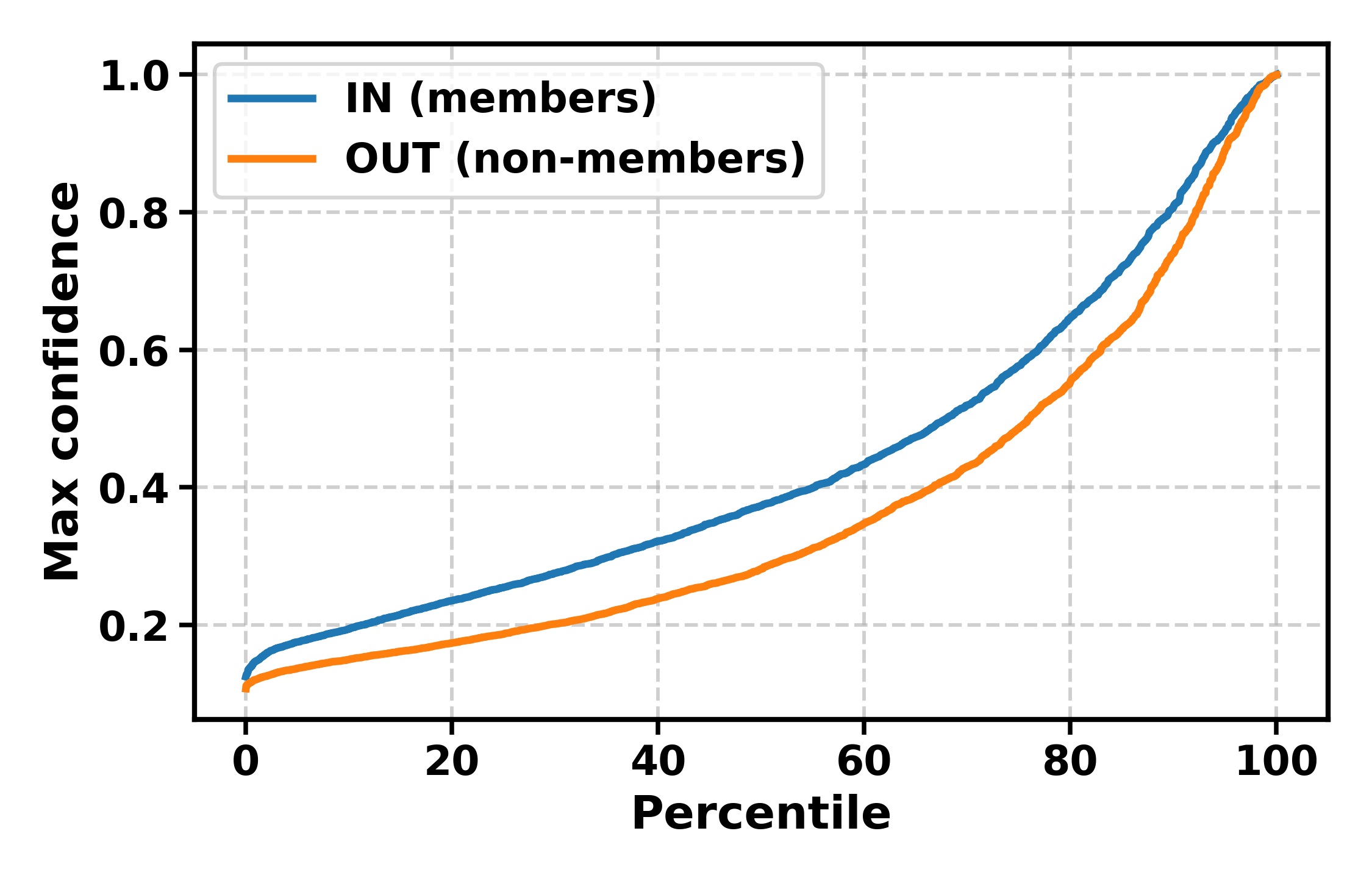}
        \smallskip

        \small (a) Maximum confidence before (left) and after (right) DynaNoise
    \end{minipage}
    \hfill
    \begin{minipage}{0.48\linewidth}
        \centering
        \includegraphics[width=0.48\linewidth]{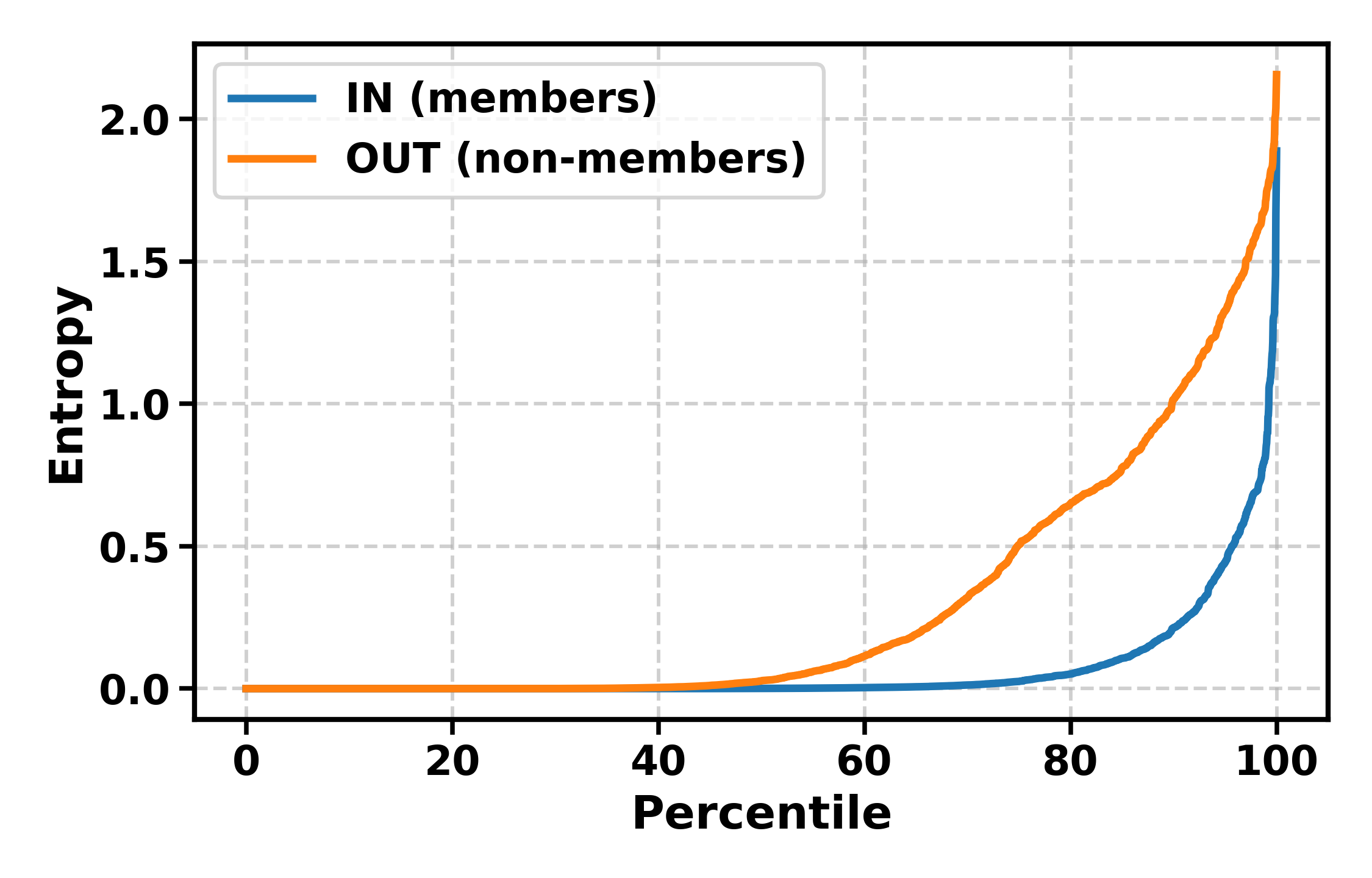}
        \includegraphics[width=0.48\linewidth]{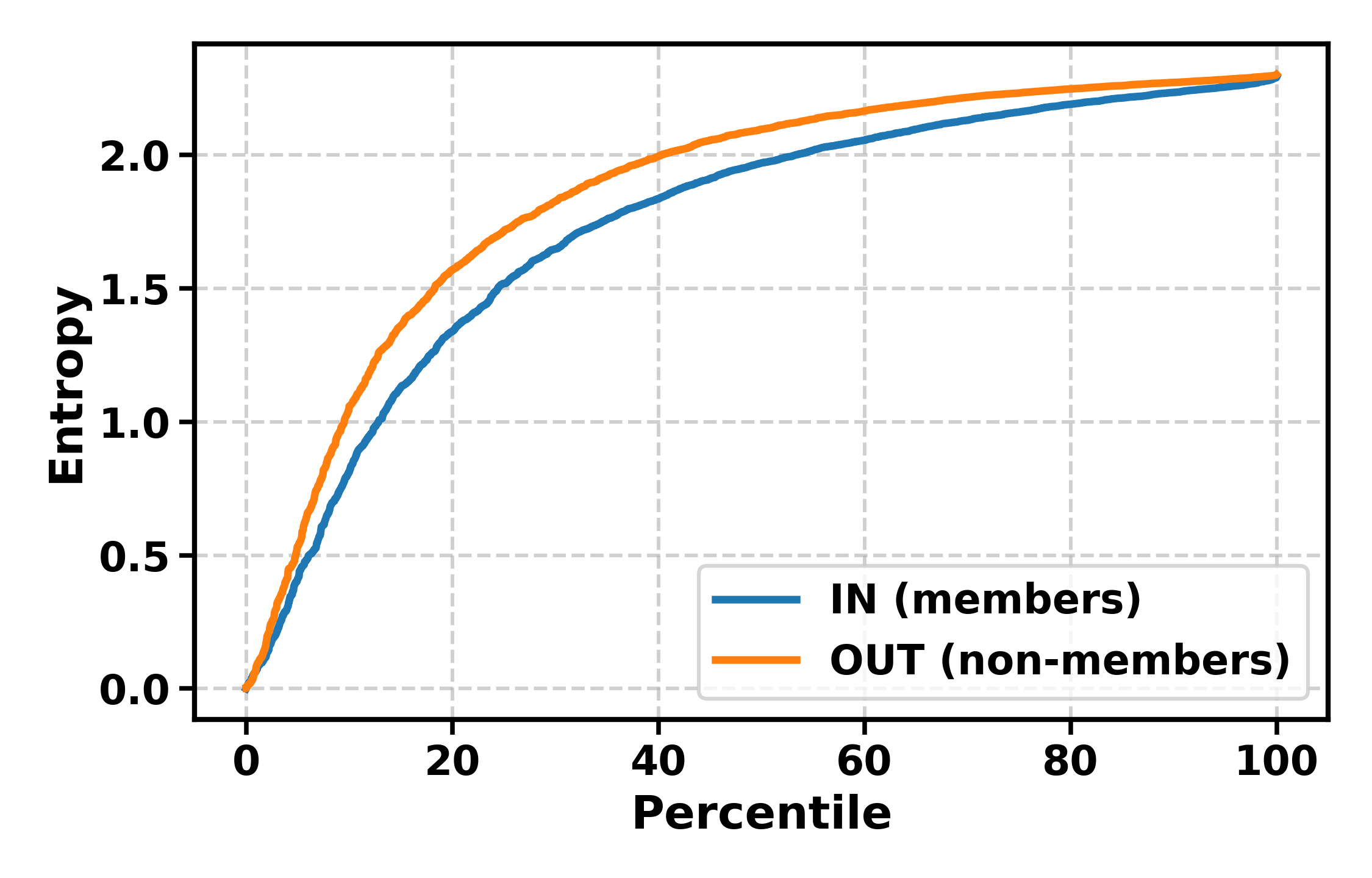}
        \smallskip

        \small (b) Predictive entropy before (left) and after (right) DynaNoise
    \end{minipage}

    \caption{CIFAR-10 percentile plots comparing confidence- and entropy-based
    statistics before and after applying DynaNoise.}
    \label{fig:cifar10_conf_entropy}
\end{figure*}

\begin{figure*}[t]
    \centering
    \begin{minipage}{0.48\linewidth}
        \centering
        \includegraphics[width=0.48\linewidth]{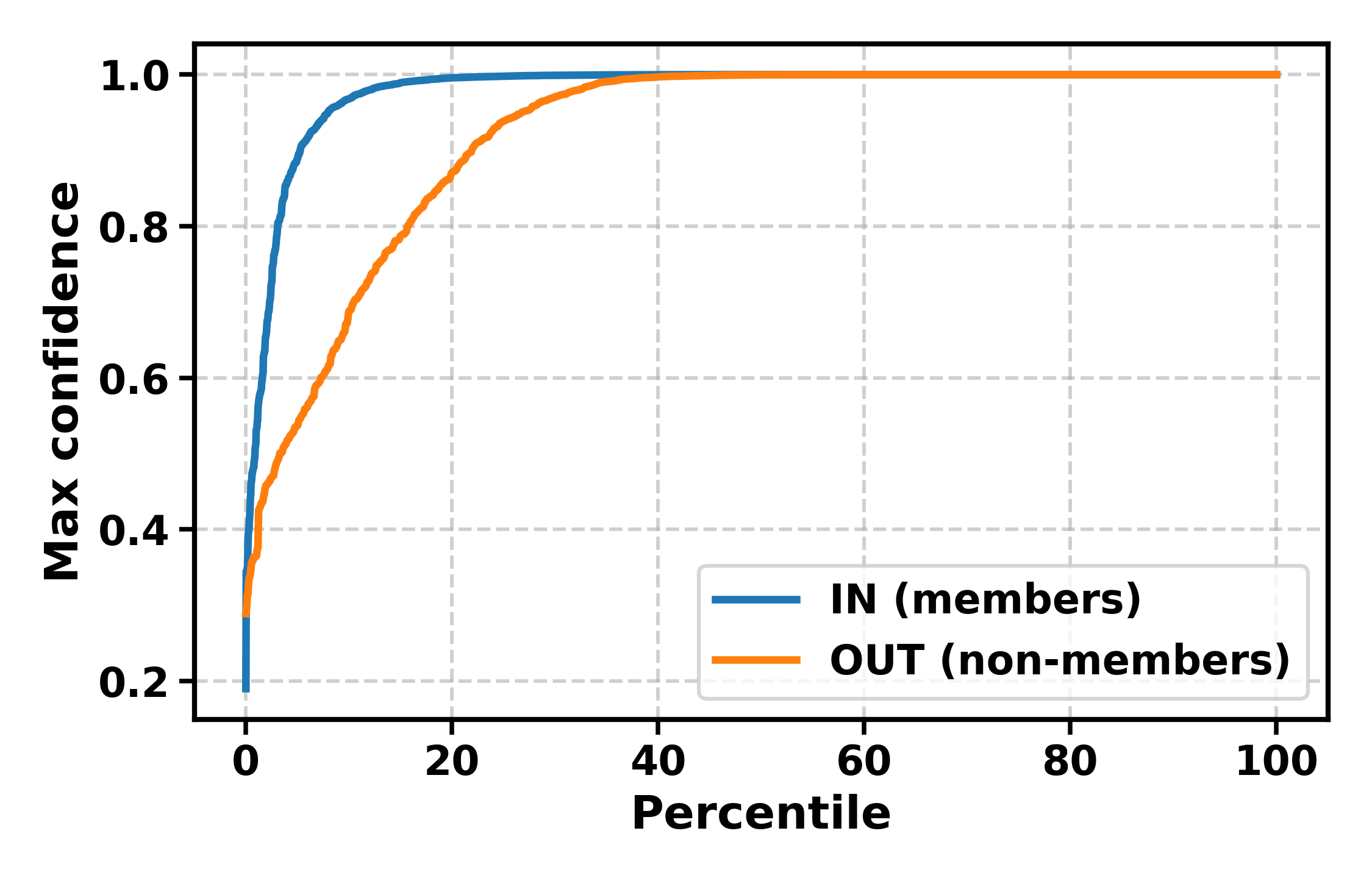}
        \includegraphics[width=0.48\linewidth]{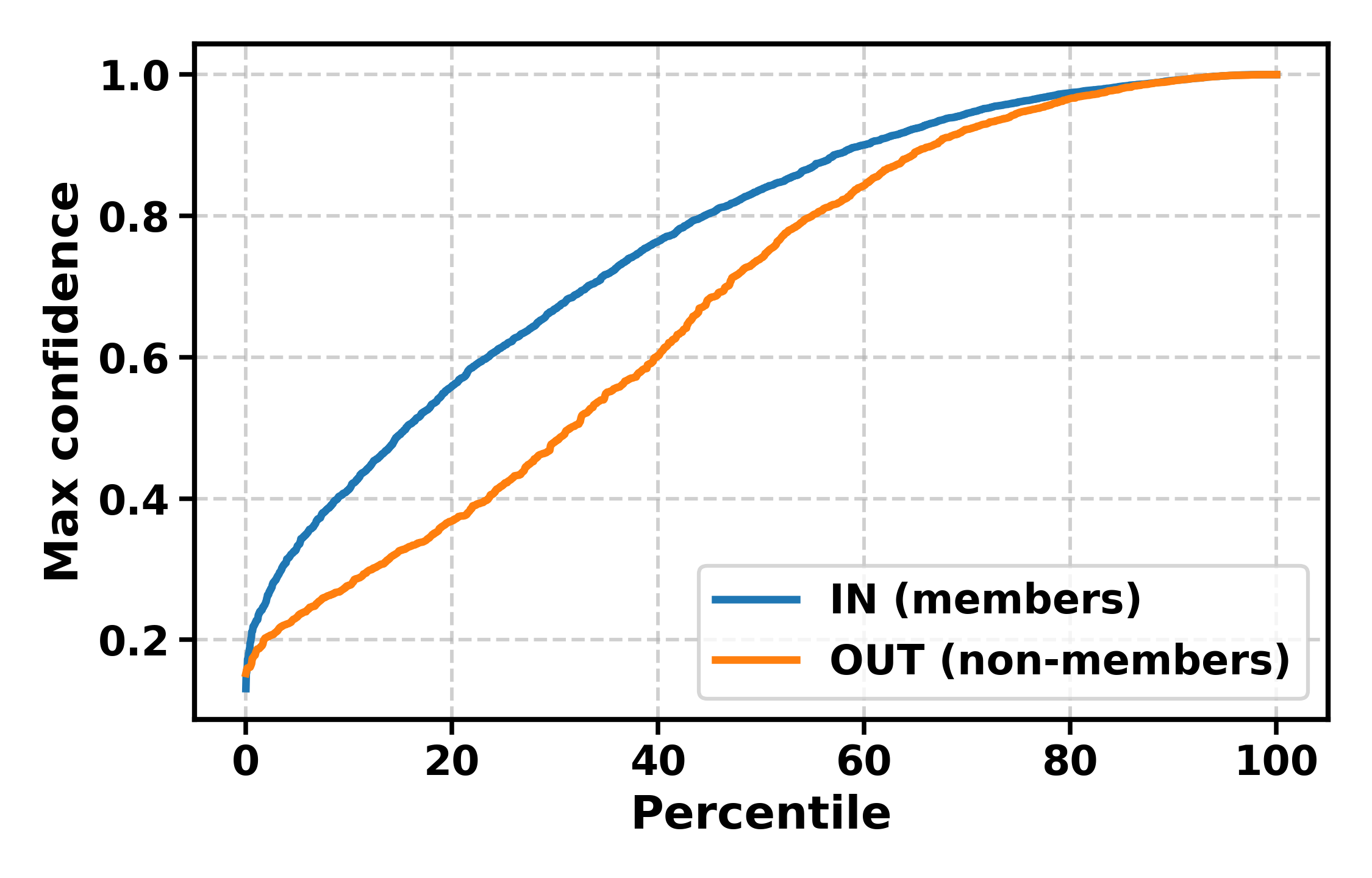}
        \smallskip

        \small (a) Maximum confidence before (left) and after (right) DynaNoise
    \end{minipage}
    \hfill
    \begin{minipage}{0.48\linewidth}
        \centering
        \includegraphics[width=0.48\linewidth]{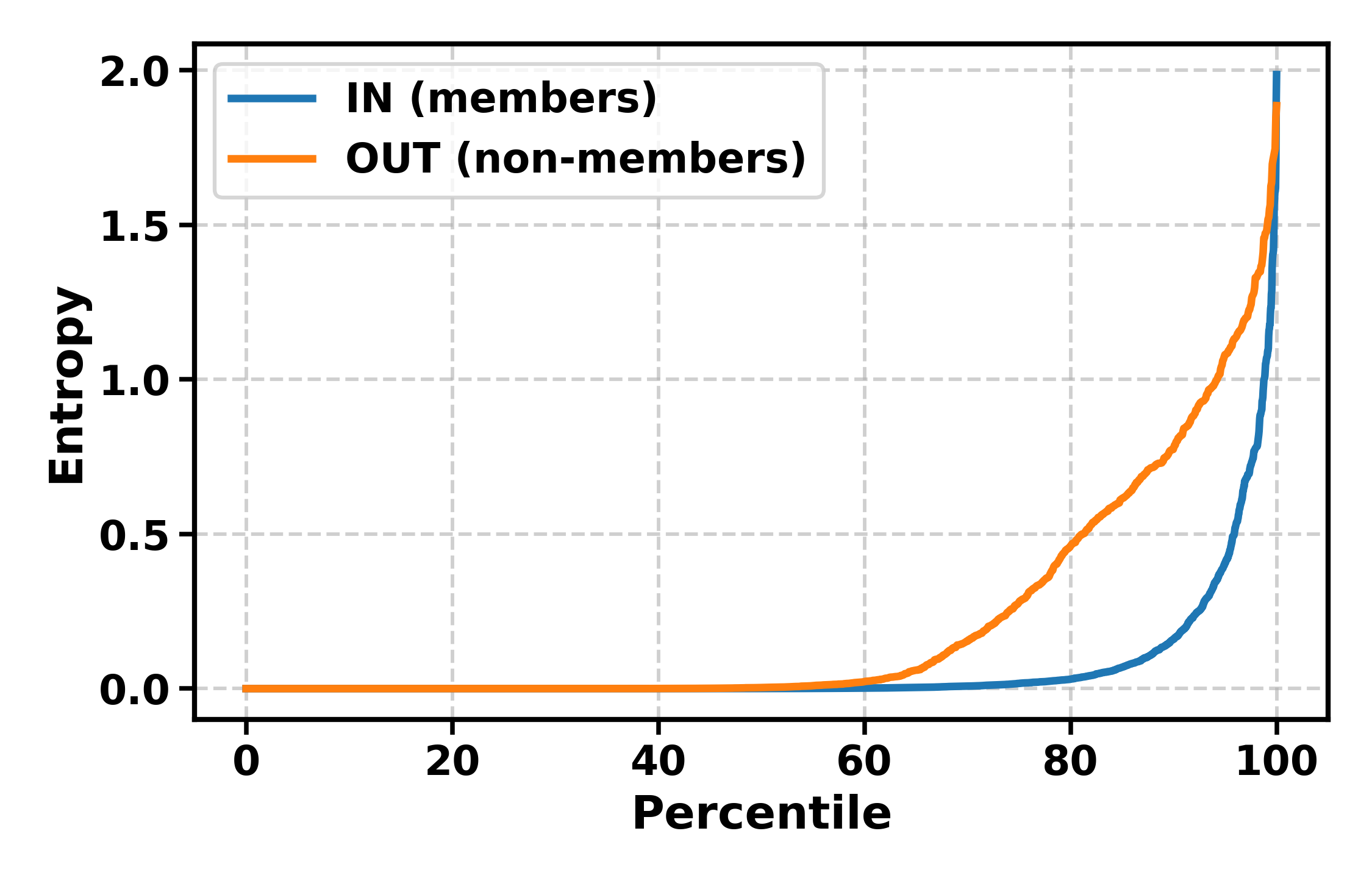}
        \includegraphics[width=0.48\linewidth]{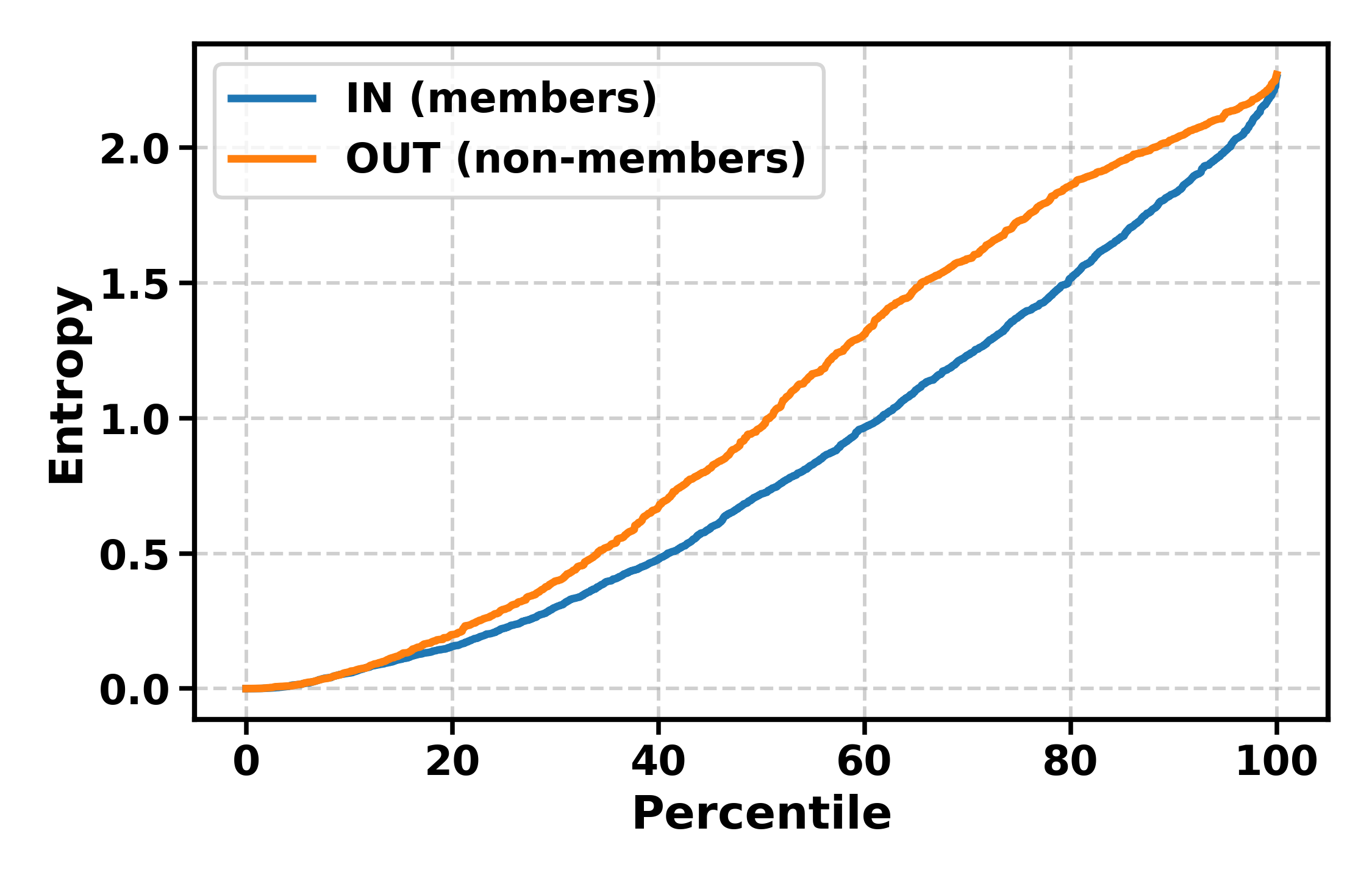}
        \smallskip

        \small (b) Predictive entropy before (left) and after (right) DynaNoise
    \end{minipage}

    \caption{ImageNet-10 percentile plots comparing confidence- and entropy-based
    statistics before and after applying DynaNoise.}
    \label{fig:imagenet10_conf_entropy}
\end{figure*}

\begin{figure*}[t]
    \centering
    \begin{minipage}{0.48\linewidth}
        \centering
        \includegraphics[width=0.48\linewidth]{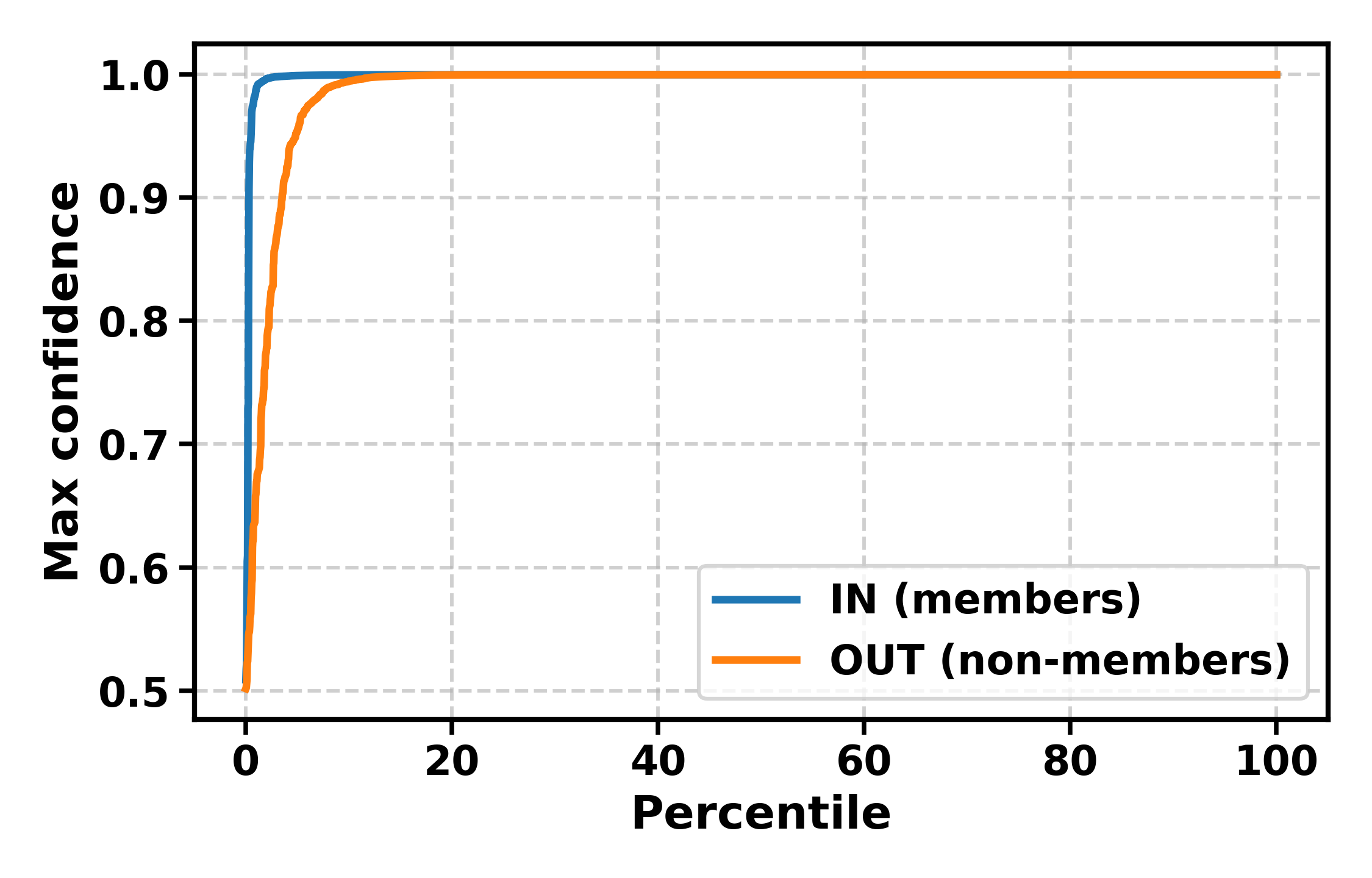}
        \includegraphics[width=0.48\linewidth]{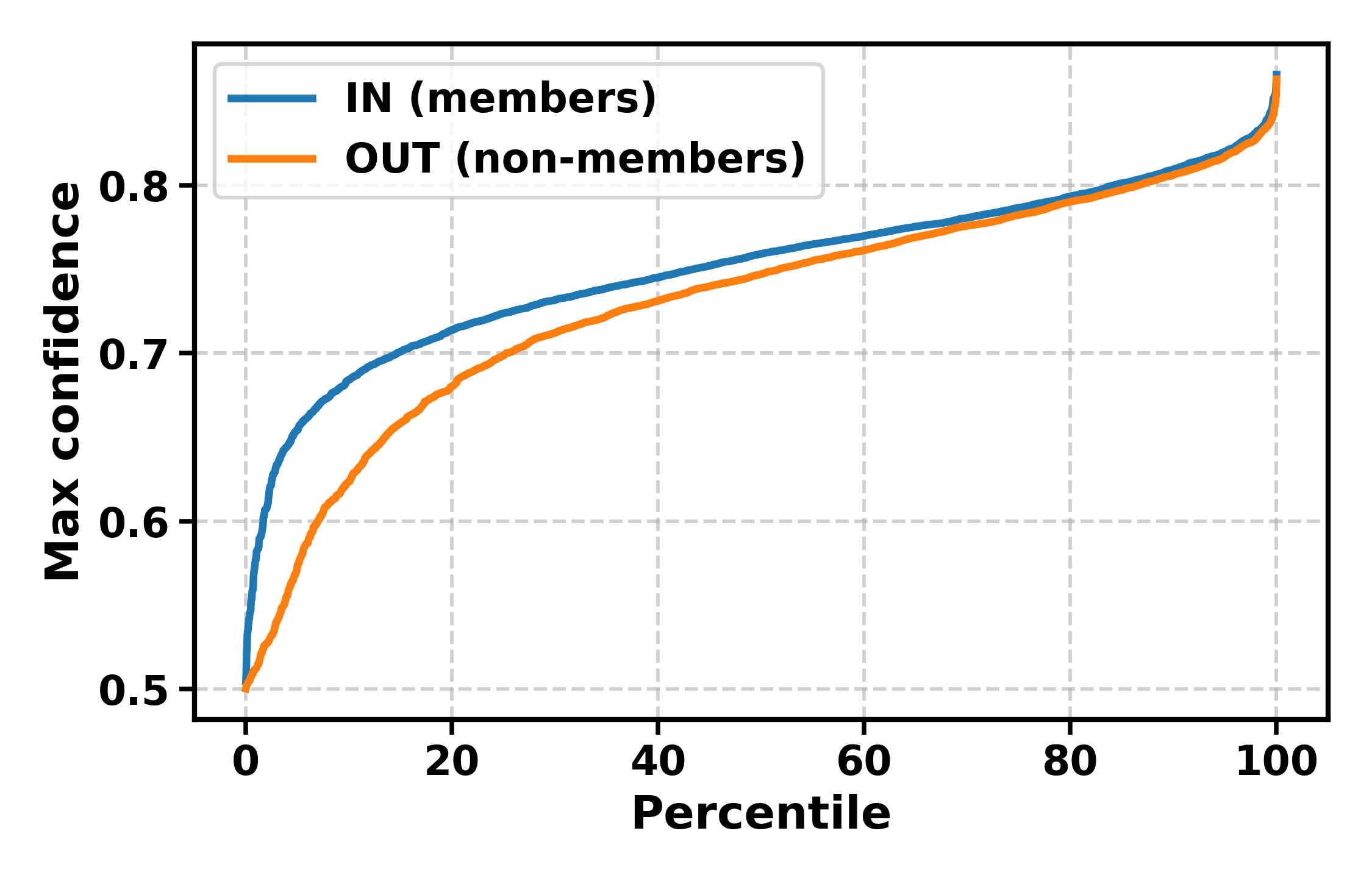}
        \smallskip

        \small (a) Maximum confidence before (left) and after (right) DynaNoise
    \end{minipage}
    \hfill
    \begin{minipage}{0.48\linewidth}
        \centering
        \includegraphics[width=0.48\linewidth]{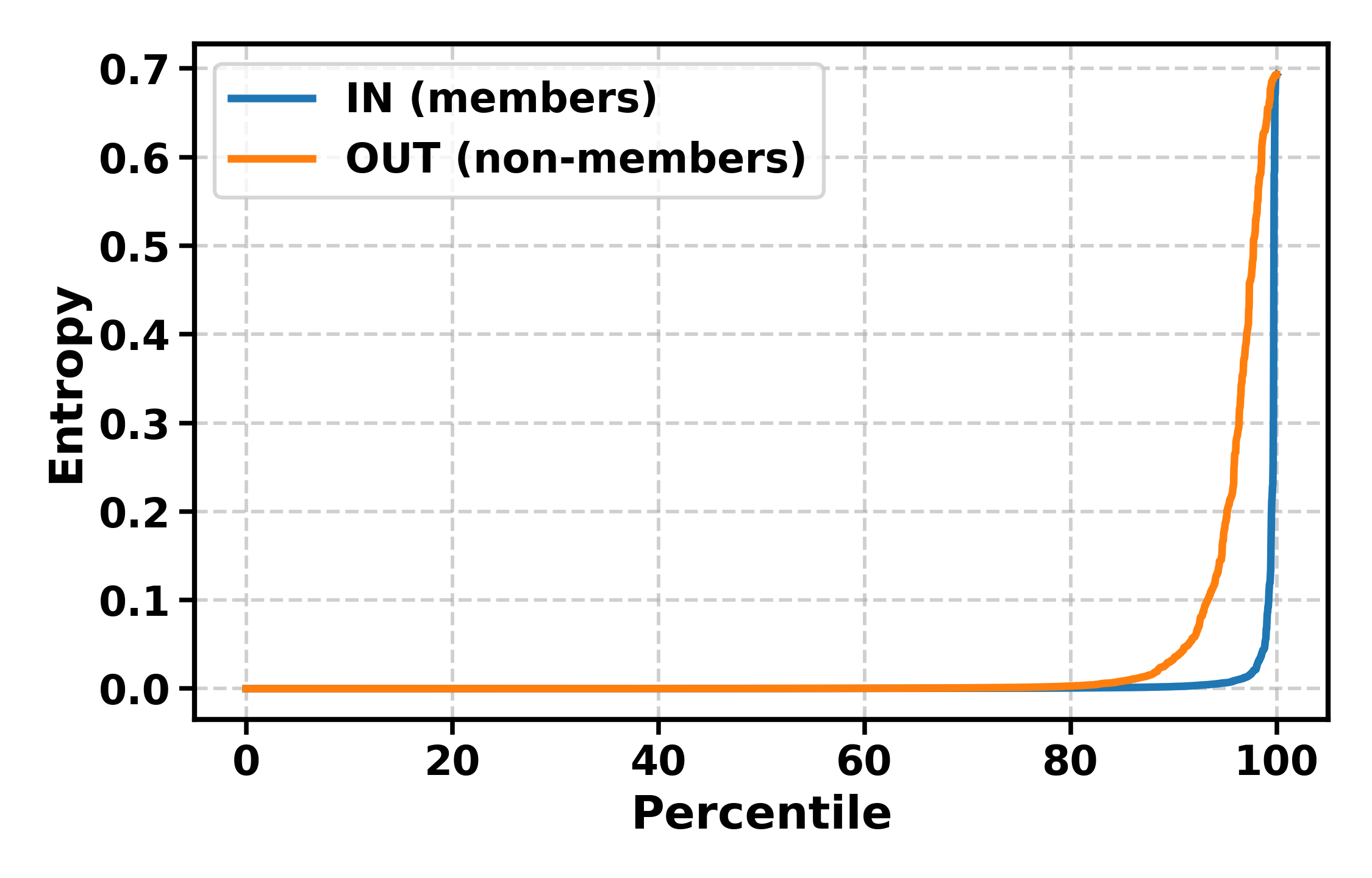}
        \includegraphics[width=0.48\linewidth]{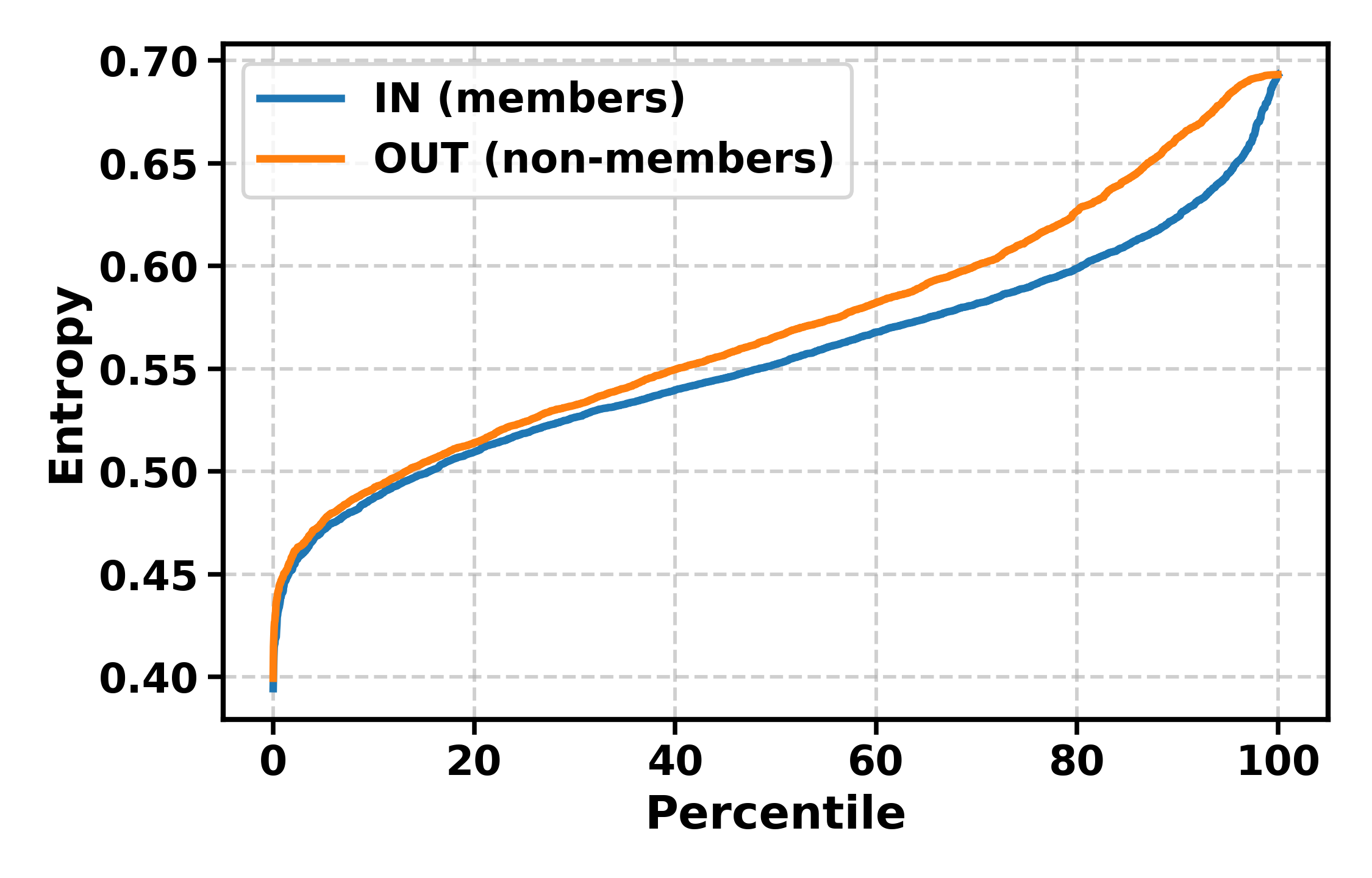}
        \smallskip

        \small (b) Predictive entropy before (left) and after (right) DynaNoise
    \end{minipage}

    \caption{SST-2 percentile plots comparing confidence- and entropy-based
    statistics before and after applying DynaNoise.}
    \label{fig:sst2_conf_entropy}
\end{figure*}

To better understand the dataset-dependent behavior of membership inference
attacks (MIAs), we analyze the distributions of \emph{maximum prediction
confidence} and \emph{predictive entropy} for training (IN) and non-training
(OUT) samples across all evaluated datasets. These statistics correspond
directly to the decision signals exploited by confidence-, entropy-, and
modified-entropy-based MIAs, and provide insight into both the source of
membership leakage and the mechanisms by which DynaNoise mitigates it. The
resulting percentile plots for CIFAR-10, ImageNet-10, and SST-2 are shown in
Figures~\ref{fig:cifar10_conf_entropy}, \ref{fig:imagenet10_conf_entropy}, and
\ref{fig:sst2_conf_entropy}, respectively.

\subsection{Methodology}
For each dataset, we compute the maximum softmax probability and the Shannon entropy of the model’s predicted class distribution for every query. We report percentile curves for IN and OUT samples separately, both \emph{before} applying any defense and \emph{after} applying DynaNoise. These curves reveal how rapidly confident or low-entropy predictions emerge for training samples relative to non-members, and how this separability changes under adaptive noise injection.

\subsection{CIFAR-10}
On CIFAR-10, we observe a moderate but consistent separation between IN and OUT samples in both confidence and entropy. Training points tend to reach higher confidence and lower entropy earlier than non-members, although the gap increases gradually rather than abruptly. This behavior explains why metric-based MIAs achieve non-trivial but limited success on this dataset.

After applying DynaNoise, the confidence and entropy gaps between IN and OUT samples are noticeably compressed across the percentile range. Importantly, this compression occurs without collapsing the overall prediction structure, consistent with the minimal utility degradation observed in Table~\ref{tab:reported_attacks}. These results illustrate that DynaNoise reduces excessive confidence disparities while preserving the underlying ranking of predictions.

\subsection{ImageNet-10}
ImageNet-10 exhibits the most pronounced IN/OUT separation among the evaluated datasets. Before applying any defense, training samples attain near-unit confidence and near-zero entropy at substantially lower percentiles than non-members, indicating a highly peaked prediction regime for memorized examples. This sharp divergence explains the strong performance of confidence-, entropy-, and LiRA-based MIAs on this dataset.

After applying DynaNoise, both confidence and entropy curves show a substantial reduction in IN/OUT separation. The early-percentile confidence spike for training samples is strongly suppressed, and entropy distributions for members and non-members become significantly more aligned across most of the percentile range. This directly explains the observed reduction in attack success rates under strong attacks and highlights the advantage of query-adaptive, entropy-scaled noise over global confidence suppression.

\subsection{SST-2}
In contrast to the vision datasets, SST-2 shows minimal separation between IN and OUT samples in both confidence and entropy even before applying defenses. Predictions remain relatively well-calibrated across percentiles, reflecting the behavior of pretrained transformer-based language models. Consequently, metric-based MIAs are intrinsically weaker on SST-2.

Applying DynaNoise results in only minor changes to the confidence and entropy distributions, further indicating that the method adapts naturally to low-risk regimes by avoiding unnecessary perturbations. This behavior aligns with the strong utility preservation observed for SST-2 while still providing protection against high-risk, low-entropy queries.

\subsection{Discussion}
Across all datasets, these results demonstrate that membership leakage is fundamentally shaped by dataset complexity and model calibration. ImageNet-10’s low-entropy, high-confidence prediction regime creates favorable conditions for metric-based MIAs, whereas SST-2’s calibrated outputs inherently limit attack effectiveness. By scaling noise injection to per-query entropy, DynaNoise adapts naturally to these dataset-specific characteristics, reducing confidence-based separability where it is most pronounced while preserving utility in lower-risk regimes.

\section{Additional Computational Overhead Results}
\label{app:overhead}

\subsection{Empirical Training and Inference Overhead}

Table~\ref{tab:overhead} reports empirical measurements of training and inference
overhead for all evaluated defenses. Inference latency is measured as the
end-to-end per-sample runtime at deployment, including any defense-specific
post-processing applied to model outputs. Inference overhead is reported
relative to the undefended baseline model. Training overhead is reported as the ratio between the total training cost of a
defended model and the corresponding undefended baseline.
\begin{table*}[t]
\centering
\caption{Empirical computational overhead of evaluated defenses. Inference
latency is measured in milliseconds per sample. Training overhead is reported relative to the undefended baseline.}
\label{tab:overhead}
\begin{tabular}{lcccc}
\toprule
\textbf{Defense} &
\textbf{End-to-End Inference} &
\textbf{Inference Overhead} &
\textbf{Training Overhead} &
\textbf{Retraining Required} \\
& \textbf{(ms/sample)} & \textbf{($\times$ baseline)} & \textbf{($\times$ baseline)} & \\
\midrule
None        & 0.1414 & 1.0  & 1.0  & No  \\
AdvReg      & 0.1414 & 1.0  & 1.0  & Yes \\
MemGuard    & 28.1118 & 198.8 & 1.66 & No  \\
RelaxLoss   & 0.1414 & 1.0  & 1.0  & Yes \\
SELENA      & 0.1414 & 1.0  & 2.61 & Yes \\
HAMP        & 0.2819 & 2.0  & 1.0  & No  \\
\textbf{DynaNoise}   & 0.1432 & 1.01 & 1.0  & No  \\
\bottomrule
\end{tabular}
\end{table*}

\subsection{Overhead Definitions}

Let $T_{\text{model}}$ denote the per-sample forward-pass latency of the
undefended target model, and let $T_{\text{defense}}$ denote any additional
defense-specific computation performed at inference time. The end-to-end
inference latency is defined as
\begin{equation}
T_{\text{e2e}} = \frac{1}{N} \sum_{i=1}^{N}
\bigl(T_{\text{model}}^{(i)} + T_{\text{defense}}^{(i)}\bigr),
\end{equation}
where $N$ is the number of evaluated samples.

Inference overhead is reported as a multiplicative ratio relative to the
undefended baseline:
\begin{equation}
\text{Inference Overhead} =
\frac{T_{\text{e2e}}^{\text{defended}}}
     {T_{\text{e2e}}^{\text{baseline}}}.
\end{equation}

Training overhead is defined as the ratio between the total training cost of a
defended model and that of the corresponding undefended baseline.
\begin{equation}
\text{Training Overhead} =
\frac{T_{\text{train}}^{\text{defended}}}
     {T_{\text{train}}^{\text{baseline}}}.
\end{equation}

Here, $T_{\text{train}}^{\text{defended}}$ includes all training necessary for
deployment, which may consist of retraining the target model (for training-time
defenses) or training auxiliary components (e.g., defense classifiers), but does
not include baseline models trained solely for experimental comparison.

\subsection{Discussion}

Post-hoc defenses exhibit markedly different computational profiles. MemGuard
incurs extreme inference-time overhead due to per-query iterative optimization,
resulting in orders-of-magnitude slower inference. HAMP approximately doubles
inference latency due to an additional forward evaluation and output
manipulation step. Training-time defenses such as AdvReg, RelaxLoss, and SELENA
require retraining the target model; among these, SELENA exhibits substantially
higher training cost due to training multiple sub-models and a subsequent
distillation phase, while AdvReg and RelaxLoss have training costs comparable to the corresponding undefended baseline.

In contrast, DynaNoise operates entirely post hoc, requires no retraining or
auxiliary models, and adds only a small ($\approx$1\%) inference-time overhead
corresponding to a single noise injection and softmax operation. As a result,
DynaNoise achieves strong privacy protection with the smallest combined training
and inference overhead among all evaluated defenses, making it particularly well
suited for deployment on pretrained models in large-scale or resource-constrained
settings.

\FloatBarrier

\end{document}